\documentclass[letterpaper,pra,superscriptaddress,showpacs,floats,twocolumn,ps2pdf]{revtex4}
\usepackage[latin1]{inputenc}
\usepackage[T1]{fontenc}
\usepackage{placeins}
\usepackage{amsmath,amssymb}

\usepackage{graphicx,color}
\usepackage{hyperref}
\usepackage{hypernat}

\definecolor{rltred}{rgb}{0.75,0,0}
\definecolor{rltgreen}{rgb}{0,0.5,0}
\definecolor{rltblue}{rgb}{0,0,0.75}
\hypersetup{colorlinks,%
hypertexnames=true,%
pdfauthor={Johannes Feist, Stefan Nagele, Renate Pazourek, Emil Persson, Barry I. Schneider, Lee A. Collins, Joachim Burgd\"orfer},%
pdftitle={Nonsequential two-photon double ionization of helium},%
pdfview=FitH,%
pdfstartview=FitH,%
pdfpagemode=UseNone,%
bookmarksopen=true,%
bookmarksnumbered=true,%
pdfhighlight=/I,%
linkcolor=rltred,%
citecolor=rltgreen,%
linktocpage=true}

\begin{document}
\newcommand{\Int}{\int\limits}
\newcommand{\IInt}{\iint\limits}
\newcommand{\IIInt}{\iiint\limits}
\newcommand{\IIIInt}{\iiiint\limits}

\newcommand{\etal}{\emph{et~al.\@} }
\newcommand{\ie}{i.e., }
\newcommand{\Schro}{Schr\"o\-din\-ger }
\newcommand{\eg}{e.g.\@ }
\newcommand{\cf}{cf.\@ }

\newcommand{\expval}[1]{\langle#1\rangle}
\newcommand{\abs}[1]{\left|#1\right|}
\newcommand{\norm}[1]{\left\|#1\right\|}

\newcommand{\bra}[1]{\langle#1|}
\newcommand{\ket}[1]{|#1\rangle}
\newcommand{\braket}[2]{\langle#1|#2\rangle}

\renewcommand{\equationautorefname}{Eq.}
\renewcommand{\figureautorefname}{Fig.}

\newcommand{\cvec}[1]{\mathbf{#1}}
\newcommand{\op}[1]{\mathrm{\hat{#1}}}
\newcommand{\vecop}[1]{\cvec{\hat{#1}}}
\newcommand{\eqcomma}{\,,\;\;}
\newcommand{\ed}{\,}

\newcommand{\dd}{\mathrm{d}}
\newcommand{\dx}{\dd x}
\newcommand{\dt}{\dd t}
\newcommand{\dr}{\dd r}
\newcommand{\dW}{\dd\Omega}
\newcommand{\dE}{\dd E}
\newcommand{\dk}{\dd k}
\newcommand{\Dt}{\Delta t}

\newcommand{\csph}{{\mathcal{Y}}}
\newcommand{\sph}[2]{{\mathrm{Y}_{#2}^{(#1)}}}

\newcommand{\submax}{\mathrm{max}}
\newcommand{\Lmax}{L_\submax}
\newcommand{\lonemax}{l_{1,\submax}}
\newcommand{\ltwomax}{l_{2,\submax}}

\newcommand{\cmfs}{\,\text{cm}^4\text{s}}
\newcommand{\ev}{\,\text{eV}}
\newcommand{\au}{\,\text{a.u.}}
\newcommand{\Wcm}{\,\text{W}/\text{cm}^2}
\newcommand{\as}{\,\text{as}}
\newcommand{\fs}{\,\text{fs}}

\newcommand{\Teff}{{T_{\mathrm{eff}}}}
\newcommand{\Etot}{{E_\mathrm{tot}}}

\title{Nonsequential two-photon double ionization of helium}

\author{J.~Feist}
\email{johannes.feist@tuwien.ac.at} 
\affiliation{Institute for Theoretical Physics, 
             Vienna University of Technology, 1040 Vienna, Austria, EU}

\author{S.~Nagele}
\affiliation{Institute for Theoretical Physics, 
             Vienna University of Technology, 1040 Vienna, Austria, EU}

\author{R.~Pazourek}
\affiliation{Institute for Theoretical Physics, 
             Vienna University of Technology, 1040 Vienna, Austria, EU}
             
\author{E.~Persson}
\affiliation{Institute for Theoretical Physics, 
             Vienna University of Technology, 1040 Vienna, Austria, EU}

\author{B.~I.~Schneider}
\affiliation{Physics Division, 
             National Science Foundation, Arlington, Virginia 22230, USA}
\affiliation{Electron and Atomic Physics Division, 
             National Institute of Standards and Technology, Gaithersburg, Maryland 20899, USA}

\author{L.~A.~Collins}
\affiliation{Theoretical Division, T-4,
             Los Alamos National Laboratory, Los Alamos, New Mexico 87545, USA}

\author{J.~Burgd\"orfer}
\affiliation{Institute for Theoretical Physics, 
             Vienna University of Technology, 1040 Vienna, Austria, EU}

\date{\today}

\begin{abstract}
We present accurate time-dependent ab initio calculations on fully differential and total integrated (generalized) cross sections for the nonsequential two-photon double ionization of helium at photon energies from $40$ to $54\ev$\ed. Our computational method is based on the solution of the time-dependent \Schro equation and subsequent projection of the wave function onto Coulomb waves. We compare our results with other recent calculations and discuss the emerging similarities and differences. We investigate the role of electronic correlation in the representation of the two-electron continuum states, which are used to extract the ionization yields from the fully correlated final wave function. In addition, we study the influence of the pulse length and shape on the cross sections in time-dependent calculations and address convergence issues.
\end{abstract}
\pacs{32.80.Rm, 32.80.Fb, 42.50.Hz}

\maketitle

\section{Introduction}
Double ionization of helium has long been of great interest in atomic physics  since it provides fundamental insights into the role of electronic correlation in the full three-body Coulomb break-up process.  Understanding the dynamics in this simple, two-electron system is crucial to understanding more complex atoms and even simple molecules~\cite{Byron67,Abe1970,DalSad1992,Burg92,Proulx93,Pont95,PindRob98,DunTayParSmy1999,KorLam1999,Becker99,Lein2000,Parker2000,MerHarNic2001,IshMid2005,IstProManMarSta2006,MakNikLam2001}.
With the advent of intense light sources in the vuv and xuv region~\cite{FLASH2007,Dromey06,NauNeeSok2004,Tsak06,SerYakSer2007,GibPauWag2003,ZhaLytPop2007}, the focus of interest has switched from one-photon to multi-photon processes.
Specifically, two-photon double ionization has been the subject of intense theoretical studies. Several authors  \cite{ColPin2002,LauBac2003,HuCoCo2005,FouLagEdaPir2006,FouAntPir2008,IvaKhe2007,NikLam2007,NikLam2001,FenHar2003,HorMorRes2007} have calculated generalized cross sections for the nonsequential two-photon ionization process in the energy range from $40$ to $54\ev$ using a wide variety of computational methods.   Despite  considerable efforts, quantitative agreement between the different calculations has not yet been reached. The reasons for the remaining discrepancies are the subject of ongoing discussions. In particular, there have been speculations that the representation of the double continuum might be responsible for the existing differences. Nevertheless, even for methods which take correlation into account in the final states, the cross sections obtained still disagree, and a systematic change in the results due to the improved treatment of electronic correlation has not yet been observed.
Recently, experimental data has become available as well. Hasegawa, Nabekawa \etal\cite{HasTakNabIsh2005,NabHasTakMid2005} used the $27$th harmonic at $41.8\ev$ of a femtosecond pulse from a Ti:sapphire laser, and Sorokin \etal\cite{SorWelBob2007} performed their experiment at the FLASH free-electron laser in Hamburg at $42.8\ev$. While this is a good beginning, the uncertainties in the data are too large to help in resolving differences in the theoretical calculations.
With the current rapid progress in intense xuv sources, further experiments that cover larger energy ranges can be expected in the future which might help to clarify the situation.

Calculations for two-photon ionization employ either a time-independent (TI) or a time-dependent (TD) approach. TI methods involve either lowest-order perturbation theory (LOPT) or $R$-matrix Floquet theory. TD methods are based on a direct solution of the time-dependent \Schro equation and are therefore not restricted to any given order of the perturbation.
Thus, they can be applied equally well to the strong field regime. In the present case of moderate intensities of the xuv field ($\sim\!10^{12}\Wcm$), corrections to LOPT are expected to be small. The decisive advantage of TD methods comes here from a different aspect. Namely, TI calculations of processes involving correlated two-electron final states in the continuum, $\Psi_{\cvec k_1,\cvec k_2}(\cvec r_1,\cvec r_2)$, require the knowledge of the final state in the entire configuration space in order to calculate the transition amplitude
\begin{equation}\label{eq:transamp}
t_{i\to \cvec k_1, \cvec k_2} = \bra{\Psi_{\cvec k_1,\cvec k_2}} U^{(N)} \ket{\Psi_i} \,,
\end{equation}
where $U^{(N)}$ is the transition operator for an $N$-photon process ($N=2$ in the following).
As the numerical or analytical determination of accurate correlated continuum final  states remains a challenge, evaluation of \autoref{eq:transamp} involves, inevitably, additional approximations that are difficult to control. Adding the time as an additional degree of freedom to the six spatial dimensions of the two-electron problem allows one to bypass the determination of $\Psi_{\cvec k_1,\cvec k_2}$.
Propagating the wave packet for sufficiently long times enables the extraction of the relevant dynamical information entirely from the asymptotic region where electron correlations become negligible. Moreover, residual errors can be controlled by systematically varying the propagation time. This advantage comes along with a distinct disadvantage: Results will, in general, depend on the time-structure imposed on the external perturbation, specifically on the duration and temporal shape of the xuv pulse. A comparison with TI calculations on the level of (generalized) cross sections therefore requires a careful extraction of information and checks of the independence from pulse parameters.

In our theoretical approach, we solve the time-dependent \Schro equation (TDSE) using the time-dependent close-coupling (TDCC) scheme, \cf \cite{ColPin2002,HuCoCo2005,LauBac2003,PinRobLoc2007}. For the spatial discretization, we employ a finite element discrete variable representation (FEDVR), and the temporal propagation is performed by means of the short iterative Lanczos (SIL) procedure with adaptive time-step control. 
We present detailed convergence tests as a function of pulse duration, pulse shape, duration of propagation, gauge, spatial grid structure, and partial wave decomposition. For two-photon double ionization in the photon energy range $40$ to $50\ev$, we reach an accuracy on the $2\%$ level. Our present results are compared with available experimental and theoretical data. Atomic units are used unless indicated otherwise.

\section{Method of propagation}

The interaction of a helium atom (with infinite nuclear mass) with linearly polarized light is described by the Hamiltonian
\begin{equation}\label{eq:HeHam}
\op{H}=\frac{\vecop{p}_1^2}{2}+\frac{\vecop{p}_2^2}{2}
-\frac{2}{\op{r}_1}-\frac{2}{\op{r}_2}
+\frac{1}{\abs{\vecop{r}_1-\vecop{r}_2}}
+\op{H}_\textnormal{em}^{l,v} \,,
\end{equation}
where the interaction with the electromagnetic field in the dipole approximation is either given in \emph{length gauge} by 
\begin{equation}
\op{H}_\textnormal{em}^{l}=E(t)(\op{z}_1+\op{z}_2)
\end{equation}
or in \emph{velocity gauge} by
\begin{equation}
\op{H}_\textnormal{em}^{v}=\frac{A(t)}{c}\left(\op{p}_{z,1}+\op{p}_{z,2}\right)+\frac{A^2(t)}{c^2} \ed.
\end{equation}
If the exact solution were available, the two gauges would be strictly equivalent. Within approximate solutions, however, discrepancies may arise. The degree of gauge dependence can therefore be exploited as a measure for the convergence of the numerical solution toward the exact solution.

\subsection{Time-dependent close-coupling}

In order to solve the time-dependent \Schro equation 
\begin{equation}\label{eq:SchroCoord}
i\frac{\partial}{\partial t}\Psi(\cvec{r}_1,\cvec{r}_2,t) = \op{H}\Psi(\cvec{r}_1,\cvec{r}_2,t) \,,
\end{equation}
we expand the six-dimensional wave function $\Psi(\cvec{r}_1, \cvec{r}_2)$ in coupled spherical harmonics
\begin{equation}\label{eq:pwe}
\Psi(\cvec{r}_1,\cvec{r}_2,t)=
\sum_{L,M}^{\infty}\sum_{l_1,l_2}^{\infty}
\frac{R_{l_1,l_2}^{LM}(r_1,r_2,t)}{r_1 r_2}\csph_{l_1,l_2}^{LM}(\Omega_1,\Omega_2)
\end{equation}
with
\begin{multline}
\csph_{l_1,l_2}^{LM}(\Omega_1,\Omega_2)=\\
\sum_{m_1,m_2}
\braket{l_1 m_1 l_2 m_2}{l_1 l_2 L M}\sph{l_1}{m_1}(\Omega_1)\sph{l_2}{m_2}(\Omega_2)
\ed.
\end{multline}

Substitution of \autoref{eq:pwe} into \autoref{eq:SchroCoord} yields a system of coupled partial differential equations in $(r_1, r_2, t)$\,, the TDCC equations \cite{PinRobLoc2007}
\begin{multline}\label{eq:coupledSE}
\sum_{L,M}^{\infty}\sum_{l_1,l_2}^{\infty}\bra{l_1' l_2' L' M'}\op{H}\ket{l_1 l_2 L M}
\frac{R_{l_1,l_2}^{LM}(r_1,r_2,t)}{r_1 r_2} = \\
i\frac{\partial}{\partial t}\frac{R_{l_1',l_2'}^{L'M'}(r_1,r_2,t)}{r_1 r_2}
\eqcomma
\end{multline} 
where in practice the sums have to be truncated at certain maximum angular momenta $(\Lmax,\lonemax,\ltwomax)$\ed.

We restrict ourselves to partial waves with total $M=0$ since $M$ is conserved for linearly polarized laser fields. The helium singlet ($S=0$) ground state is space-symmetric. As the spin quantum number $S$ is conserved in the dipole approximation, the wave function is space-symmetric for all times, implying that $R_{l_2,l_1}^{L}(r_2,r_1,t)=R_{l_1,l_2}^{L}(r_1,r_2,t)$\ed. As a consequence, the functions $R_{l_1,l_2}^{L}(r_1,r_2,t)$ need not be stored for $l_2>l_1$. Together with the parity selection rules, which only allow certain combinations $(L,l_1,l_2)$, this greatly reduces the numerical effort for solving the close-coupling equations.

\subsection{Spatial discretization}

For the discretization of the radial functions $R_{l_1,l_2}^{L}(r_1,r_2,t)$, we employ a finite element discrete variable representation (FEDVR)~\cite{ResMcc2000,MccHorRes2001,Schneider05,SchColHu2006}. We divide the radial coordinates into finite elements in each of which the functions $R_{l_1,l_2}^{L}$ are represented in a local DVR basis with a corresponding Gauss-Lobatto quadrature to ensure the continuity of the wave function at the element boundaries. This method leads to sparse matrix representations of the differential operators and to a diagonal potential matrix (within quadrature accuracy). 
Additionally, the boundary condition at $r_1=r_2=0$ can be easily fulfilled by omitting the first basis function in the first finite element. The derivative discontinuity in the partial wave expansion of the electron-electron interaction at $r_1=r_2$, on the other hand, demands special treatment to guarantee an accurate representation of the Hamiltonian in the FEDVR basis~\cite{MccBaeRes2004,MccMar2004}.

\subsection{Temporal propagation}

For the temporal propagation of the solution of the coupled equations \eqref{eq:coupledSE}, we employ the short iterative Lanczos (SIL) method \cite{ParkLight86,SmyParTay1998,Lefo90} with adaptive time-step control. The initial He ground state is obtained by relaxing an arbitrary test function in imaginary time. In the SIL method, sometimes also referred to as Arnoldi-Lanczos algorithm, the time evolution operator
\begin{equation}\label{eq:U}
\op{U}(t,t+\Dt)\simeq\exp{\left[-i\op{H}(t)\Dt\right]}
\end{equation}
is represented by an $N \times N$ matrix $\op{U}^{(N)}$ in an $N$-dimensional Krylov subspace which is formally generated by repeated action of $\op{H}$ on an initial state $\Psi(t)$ and subsequent Gram-Schmidt orthogonalization.

The Lanczos algorithm is very effective because the matrix $\op{H}^{(N)}$ is tridiagonal and can be directly obtained by use of a three-term recursion relation involving mainly matrix-vector and scalar products. 
The sparse structure of the kinetic energy matrices due to the division of coordinate space into finite elements enables efficient parallelization. In our calculations, which have primarily employed computers based on cluster architectures,  we have observed linear scaling of the computational speed up to 450 processors.
This gives us the possibility to employ pulses with comparably long durations (up to a few femtoseconds) in our simulations. In addition, extensive numerical convergence tests can be performed within reasonable time (\cf \autoref{sec:conv}).

The SIL method also allows for a convenient error control since for a given subspace order $N$ the difference of the propagated wave function $\Psi^{(N)}(t+\Dt)$ to the lower-order approximation $\Psi^{(N-1)}(t+\Dt)$ can be calculated with only little extra effort and may be used as a tolerance parameter during propagation. In our implementation, we use a Krylov subspace of fixed order (with $N=12$ for all calculations presented in this work) and an adaptive time step so that $\norm{\Psi^{(N)}(t+\Dt)-\Psi^{(N-1)}(t+\Dt)}^2$ is smaller than a given tolerance parameter (typically in the order of $10^{-20}$). Hence, we achieve a high-order approximation of the exponential function (\autoref{eq:U}), and the temporal propagation is explicitly unitary and unconditionally stable. In addition, high accuracy is guaranteed due to the automatic adjustment of the time step.

\section{Extracting dynamical information}

Subsequent to the time propagation, the information on excitation and ionization probabilities and on differential and total cross sections must be extracted from the wave packet that is represented on a grid in a finite domain in coordinate space. This is a non-trivial task as straightforward projection is, for all practical purposes, precluded. Ideally, one would project onto asymptotic eigenstates of He, including its single and double ionization continua. 
As no closed solutions for the double continuum are available, this is not directly possible. Numerical diagonalization of the Hamiltonian is not feasible either. First, the basis is too large to allow diagonalization.  Second, the inclusion of the correct asymptotic boundary conditions in a numerical solution is highly non-trivial and computationally expensive. In addition, avoiding the construction of these eigenstates is one of the significant advantages of the TD approach.
We circumvent the need for projection onto exact eigenstates by exploiting the fact that we can propagate the wave packet for long times after the conclusion of the pulse. Thereafter, the three-body Coulomb system (nucleus and two electrons) has reached (near asymptotically) large distances such that projection onto approximate energy eigenstates is possible without significant error.

The states we eventually use to represent this continuum are eigenstates of the He Hamiltonian without the electron-electron interaction $\op{H}_{12}=\abs{\op{r}_1-\op{r}_2}^{-1}$, \ie products of two Coulomb waves with $Z=2$. This amounts to assuming that the electrons are far enough apart that their influence on each other can be neglected. By the same token, products of plane waves would be equally applicable when the electrons are far from the nucleus and its Coulomb potential could therefore be neglected.
The advantage of using Coulomb waves is that orthogonality to bound states is built in.
As the singly ionized part of the wave function can to a very good approximation be written as the product of an unperturbed bound state of He$^+$ and a Coulomb wave (with effective charge $Z=1$), the projection onto products of Coulomb waves is automatically orthogonal to the singly ionized part and no additional screening of the wave function has to be performed. Neglecting the electron-electron interaction, which is purely repulsive, introduces a small energy shift in the spectrum that decreases as the distance between the electrons increases.
This effect can be controlled by varying the time of projection.

\begin{figure}[tbp]
  \includegraphics[width=0.99\linewidth]{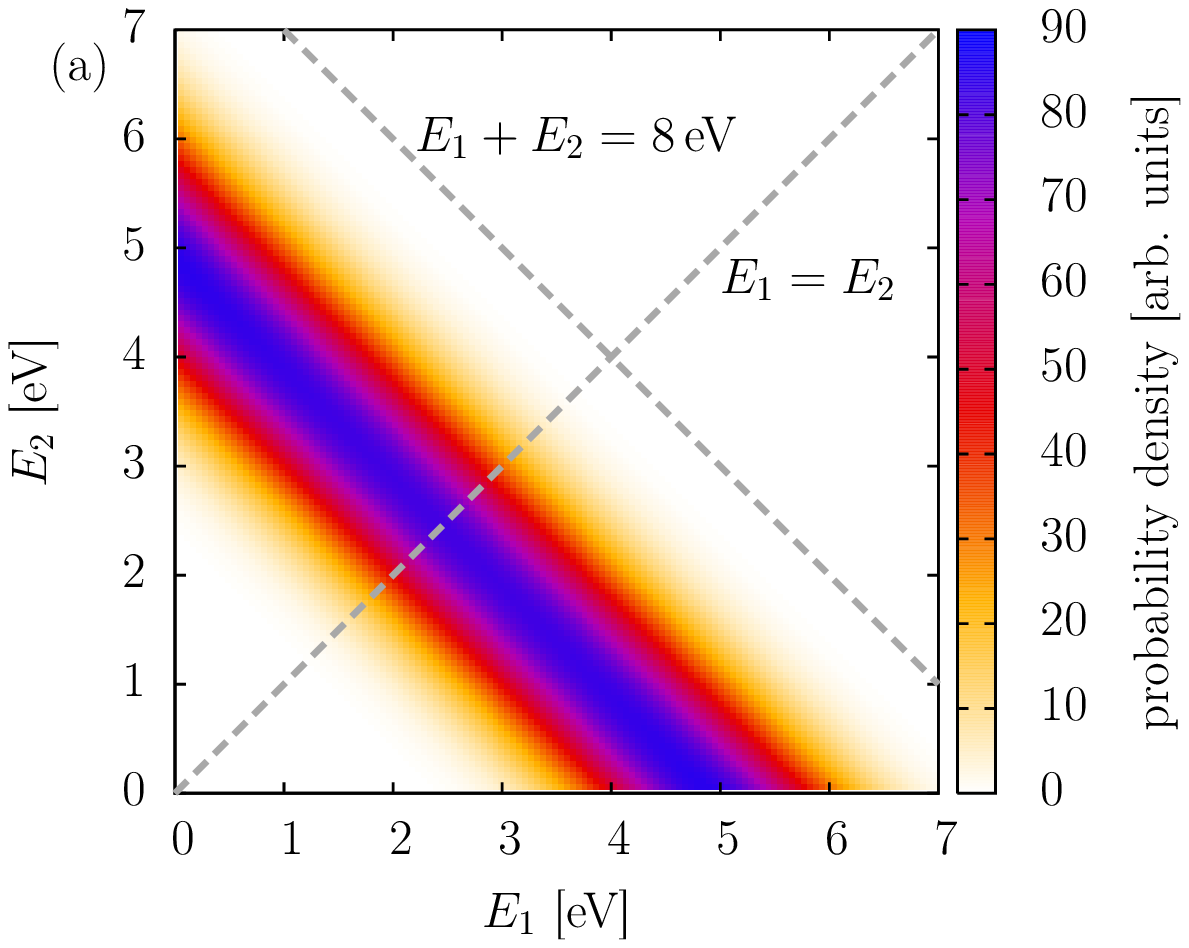}\\
  \includegraphics[width=0.99\linewidth]{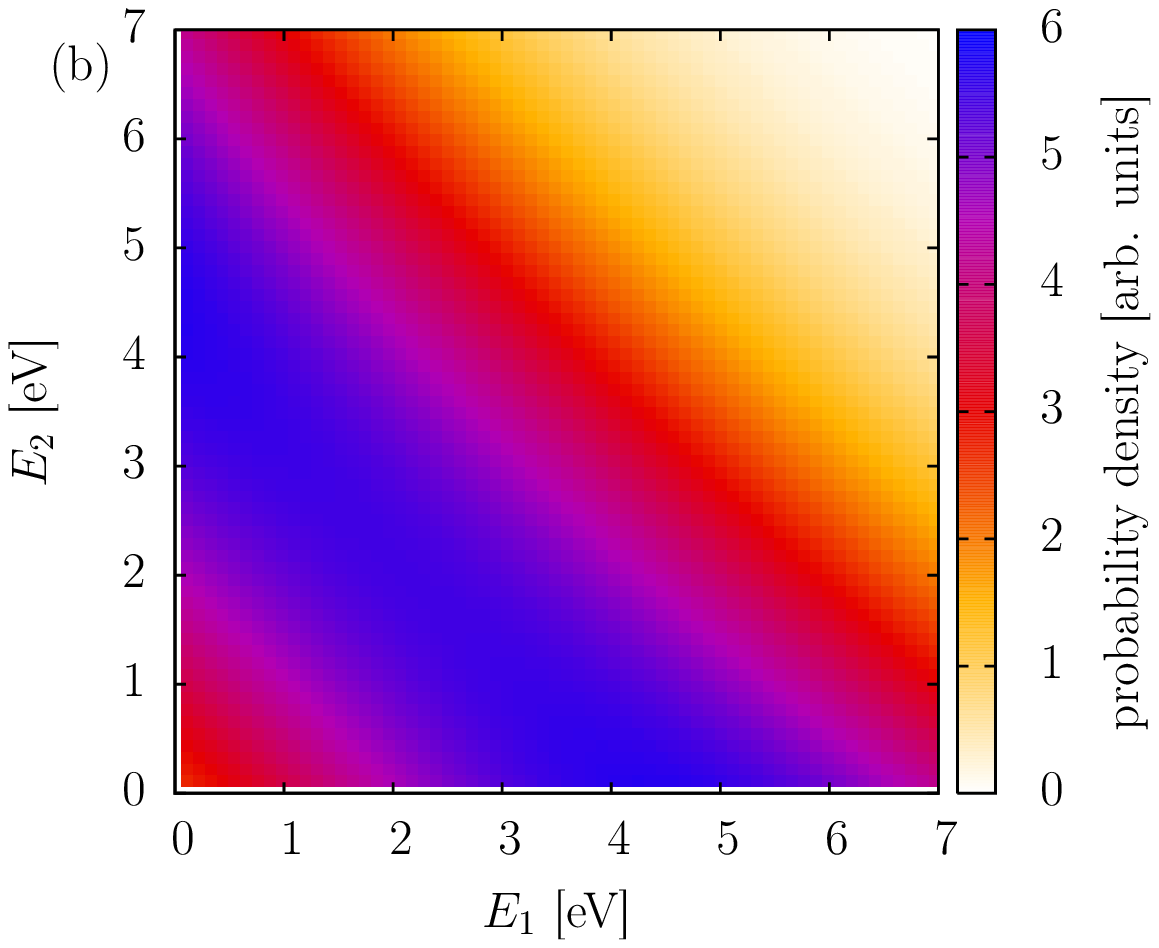}
  \caption{Energy distribution after two-photon double ionization from two different 
      laser pulses with a mean energy of $\expval{\omega}=42\ev$. The pulse (a) has a $\sin^2$ 
      envelope of total duration $4\fs$ ($\sim\!40$ cycles). The distribution is
      centered around the line $2\expval{\omega}-I_1-I_2=E_1+E_2 \approx 5\ev$, with a FWHM of about $1.5\ev$
      due to the finite duration of the pulse. The pulse (b) is a ten-cycle ($\sim\!1\fs$)
      $\sin^2$ pulse.}
  \label{fig:2dspectra}
\end{figure}

We obtain the energy-space wave function by projecting the spatial wave function onto products of energy-normalized Coulomb waves $\phi_{E,l}$, \ie
\begin{equation}\label{eq:Pkk_def}
P_{l_1,l_2}^{L}(E_1,E_2) =  \braket{\phi_{E_1,l_1}\phi_{E_2,l_2}}{R_{l_1,l_2}^{L}}.
\end{equation}
Switching from the angular momentum representation $(L,l_1,l_2)$ to the representation in angular variables $(\Omega_1,\Omega_2)$, 
the six-dimensional (effectively, five-dimensional) distribution  can be written as~\cite{ColPinRob2001,HuCoCo2005}
\begin{multline}\label{eq:Pdouble_s}
P^{DC}(E_1,E_2,\Omega_{1},\Omega_{2})=\\
\abs{\sum_{L,l_1,l_2} e^{-i\frac{\pi}{2}(l_1+l_2) + i(\sigma_{l_1}+\sigma_{l_2})}
\csph^{L0}_{l_1,l_2}(\Omega_{1},\Omega_{2}) P_{l_1,l_2}^{L}(E_1,E_2)}^2\,,
\end{multline}
where $\sigma_{l}=\arg \Gamma(1+l+i\eta)$ are the Coulomb phases.
From \autoref{eq:Pdouble_s}, reduced probability distributions can be determined by integrating over unobserved degrees of freedom.
For example, integrating over the solid angles ($\Omega_1, \Omega_2$) gives the energy distribution ($E_1,E_2$) of the electron pair (\autoref{fig:2dspectra}).

\begin{figure}[tbp]
  \includegraphics[width=\linewidth]{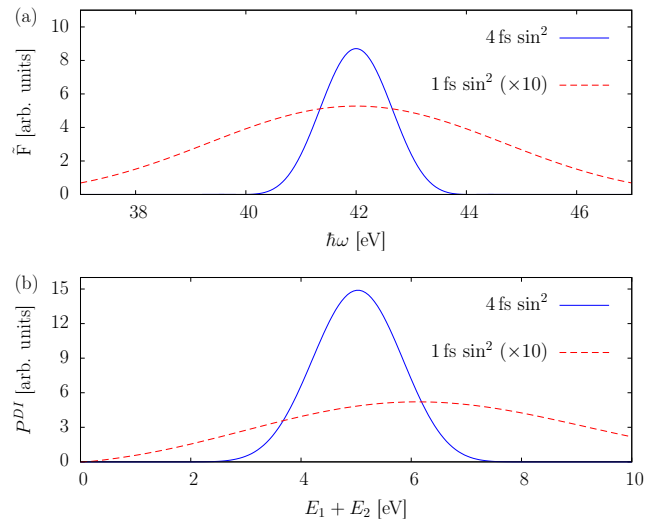}
  \caption{(a) Fourier spectra of $4\fs$ and $1\fs$ $\sin^2$ laser fields with a mean energy of $42\ev$.
      (b) Double ionization probability vs.\@ total energy $\Etot=E_1+E_2$ (\ie the integral over lines 
      with $\Etot=E_1+E_2$ from Fig.~\ref{fig:2dspectra}). For the long pulse, the 
      double ionization probability directly reflects the Fourier spectrum.
      For the shorter pulse the electron energy distribution is influenced 
      by the energy-dependence of the cross section (\cf \autoref{fig:CS}).}
  \label{fig:laser_fourier}
\end{figure}  
        
\subsection{Total cross sections}
Integrating \autoref{eq:Pdouble_s} over all variables including $E_1$ and $E_2$ gives, up to prefactors, the total double ionization cross section. The dependence on the primary photon energy is only implicit through the electromagnetic pulse entering the propagation. Within a time-dependent calculation, the resulting double ionization (DI) probability depends on the spectral distribution, \ie the shape and duration of the laser pulse, while the fundamental quantity of interest, the DI cross section (DICS) at fixed frequency of the ionizing radiation does not. Extraction of the DICS therefore requires special care.

For one-photon ionization, a straightforward relationship exists between the energy-dependent ionization yield and the cross section. From a single pulse calculation, one can thus obtain the cross section for all energies contained within the pulse~\cite{PalResMcc2008,FouLagEdaPir2006}.
This is not possible without additional approximations for two-photon or multi-photon ionization, since the relation between cross section and yield contains an integral over intermediate energies. For the evaluation of this integral, the intermediate states and energies would have to be explicitly available.
In the current approach, this is not easily possible without losing the key advantage of the time-dependent method of not having to construct intermediate or final states explicitly. 

The alternative is to use a sufficiently long pulse with narrow spectral width and calculate the cross section from the total yield with the approximation that it is constant over the width of the pulse. For this approximation to be valid, the spectral width of the pulse 
must be smaller than the energy width over which the cross section significantly changes. We can check the convergence by varying the pulse length.
Figs.~\ref{fig:2dspectra} and \ref{fig:laser_fourier} illustrate this for both the joint two-electron energy distribution $P^{DC}(E_1,E_2)$ (\autoref{fig:2dspectra}) and the integral (\autoref{fig:laser_fourier}) along lines of constant total energy $E_1+E_2$ in \autoref{fig:2dspectra} for two different pulses: a pulse with a duration of $T=4\fs$ containing about $40$ optical cycles and one with $T=1\fs$ containing about ten optical cycles. While the $4\fs$ pulse is sufficient to resolve the cross section a few eV above the threshold, the shorter pulse (frequently employed, see 
refs.~\cite{HuCoCo2005,LauBac2003,FouLagEdaPir2006,NikLam2007}) results in averaging over the threshold region.

Another requirement is that the pulse has to be weak enough such that lowest-order perturbation is applicable, and the ground state depletion can be neglected. We therefore choose a peak intensity of $I_0 = 10^{12}\Wcm$. Variation between $10^{11}\Wcm$ and $10^{13}\Wcm$ results in deviations for the total cross section at $42\ev$ of less than $0.3\%$. For an intensity of $10^{13}\Wcm$, the two-photon yield is a factor of $10^4$ higher than with $10^{11}\Wcm$. 

Another test for the applicability of perturbation theory is the linear scaling of the yield with the total duration $T$ of the pulse.
This means that the transition rate must be proportional to $\Phi(t)^N$, where $\Phi(t)=I(t)/\omega$ is the photon flux and 
$N$ is the minimum number of photons required for the process to take place. The double ionization yield is then given by
\begin{equation}\label{eq:nonseq_yield}
P^{DI}_\text{nonseq} = \Int_{-\infty}^{\infty} \dt\, \sigma_{N} \Phi(t)^N \,,
\end{equation}
where $\sigma_N$ is the total generalized $N$-photon cross section for double ionization of He.
Accordingly, the cross section is given by 
\begin{multline}\label{eq:CS_def}
 \sigma_{N} \approx \left(\frac{\omega}{I_0}\right)^N \frac{1}{\Teff_{,N}} \times\\
 \times \IIIInt\!\dE_1 \dE_2 \dW_1 \dW_2 P^{DC}(E_1,E_2,\Omega_{1},\Omega_{2})\,,
\end{multline}
where the effective time $\Teff_{,N}$ for an $N$-photon process is defined as
\begin{equation}\label{eq:teff_def}
  \Teff_{,N} = \Int_{-\infty}^{\infty}\!\dt \left(\frac{I(t)}{I_0}\right)^N \,.
\end{equation}
For a $\sin^2$ pulse envelope and a two-photon process, $\Teff_{,2}$ is found to be $35T/128$~\cite{FouLagEdaPir2006,LauBac2003,NikLam2007}.
\hyperref[eq:CS_def]{Equation \ref*{eq:CS_def}} is valid for direct, \ie nonsequential, double ionization when no on-shell intermediate state is involved.

\subsection{Differential cross sections}\label{sec:dcs}
The triply differential cross section (TDCS) for emitting one electron with energy $E_1$ into the solid angle $\Omega_1$, while the second one is emitted into $\Omega_2$, follows from \autoref{eq:CS_def} as
\begin{equation}\label{eq:tdcs_def}
 \frac{\dd \sigma_{N}}{\dE_1 \dW_1 \dW_2} = \left(\frac{\omega}{I_0}\right)^N \frac{1}{\Teff_{,N}} \int\!\dE_2 P(E_1,E_2,\Omega_1,\Omega_2) \,.
\end{equation}
The TDCS presented in this paper are all calculated in coplanar geometry, \ie $\phi_1=\phi_2=0^\circ$.
In the limit of an infinitely long laser pulse with well-defined energy (\ie a delta-like spectrum),
\autoref{eq:tdcs_def} becomes equivalent to
\begin{equation}
 \frac{\dd \sigma_{N}}{\dE_1 \dW_1 \dW_2} = \left(\frac{\omega}{I_0}\right)^N \frac{1}{\Teff_{,N}} P(E_1,N\omega-E_1,\Omega_1,\Omega_2)\,,
\end{equation}
as calculated in time-independent approaches. Unlike the joint two-electron energy distribution, the TDCS as calculated by \autoref{eq:tdcs_def} is, within reasonable limits, insensitive to the pulse shape used in the time-dependent approach since the Fourier width of the pulse is accounted for by the integration over the energy of the second electron.

Instead of specifying one of the energies and integrating over the other, it is also possible to specify energy (or momentum) sharing. For that purpose, we transform from the usual coordinates $(E_1,E_2)$ to $(\Etot, \alpha)$, with $\Etot = E_1+E_2$ and $\tan(\alpha) = E_1/E_2$. For a fixed value of $\alpha$, the integration is performed over the total energy $\Etot$, in other words, along straight lines through the origin in \autoref{fig:2dspectra}. 
This results in the TDCS at fixed energy sharing (the frequently investigated case of equal energy sharing corresponds to $\alpha=\pi/2$).

\subsection{Influence of final-state correlations}
\label{sec:final_corre}
Since the extraction of double ionization observables eventually proceeds by projection onto uncorrelated Coulomb final states, controlling and monitoring the effect of residual electron-electron correlations on the cross section becomes important. The key point is that 
electronic correlations are fully included in the initial state and in the time propagation and therefore in the wave packet at the point of projection. We monitor the residual error by 
further propagating the wave function after the conclusion of the laser pulse (\ie letting the electrons move further apart) and varying the time of projection.
If the final state were an eigenstate of the full Hamiltonian, the results would not depend on how long the projection is delayed.
The residual dependence on the time of projection is thus a direct measure of the error introduced by the neglect of final-state correlation during projection. As that time is extended, this error should become negligible. 
The maximum time one can wait is limited in practice by the box size, as the ionized wave packet will hit the box boundaries at some point and be reflected.
To test for convergence we performed one calculation with a box size of $800\au$, using the same $4\fs$ $\sin^2$ laser pulse at $42\ev$ as in \autoref{fig:2dspectra}a, and let the wave function propagate for an additional $21\fs$ after the end of the pulse. The doubly ionized part is still completely contained in the box after this time. 

\begin{figure}[tbp]
  \includegraphics[width=0.99\linewidth]{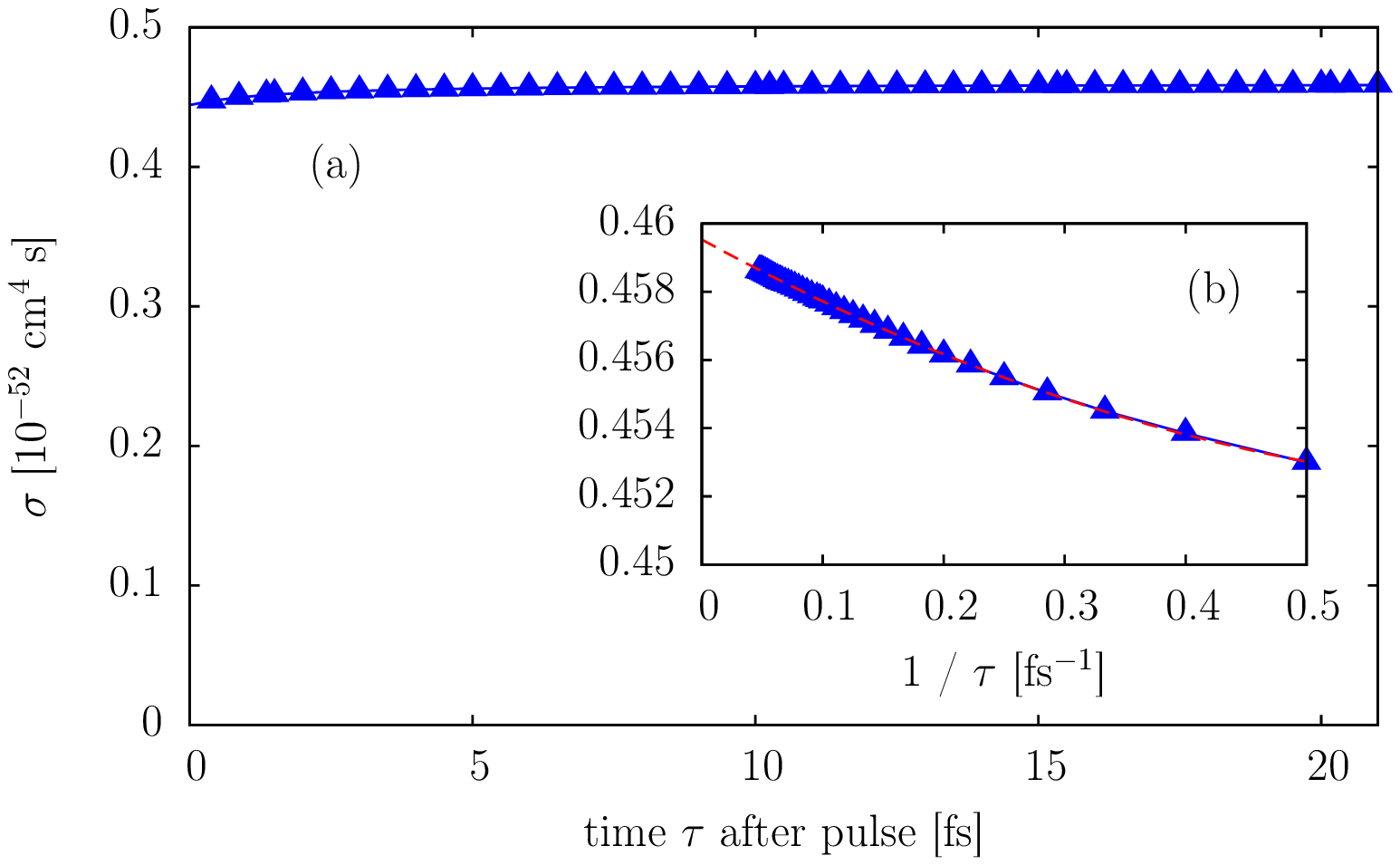}\\
  \includegraphics[width=0.99\linewidth]{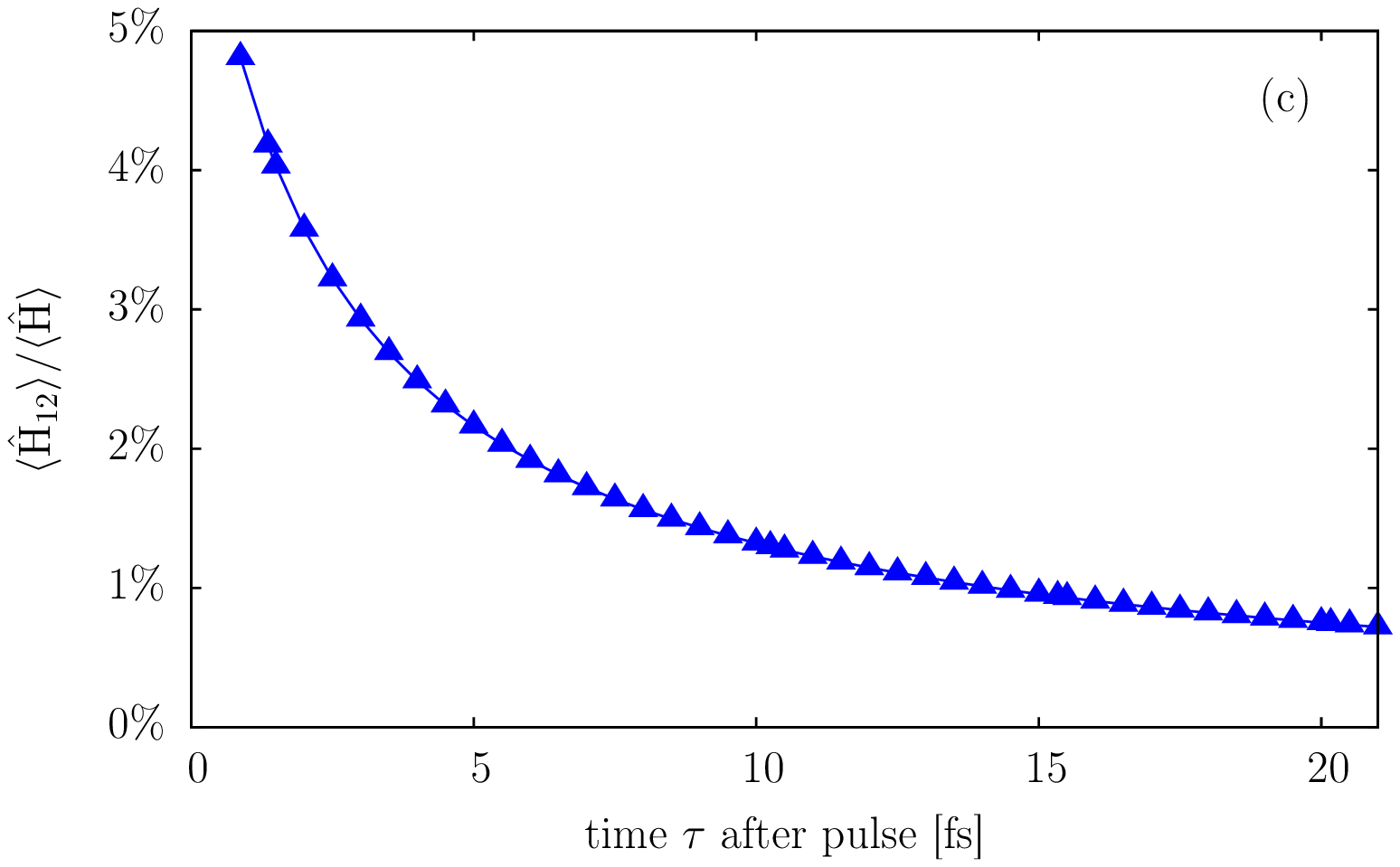}
  \caption{Convergence of the total cross section with propagation time $\tau$. The cross section is calculated at different times $\tau$ after the 
      $4\fs$ $\sin^2$ pulse from \autoref{fig:2dspectra}, with intensity $10^{12}\Wcm$ (angular momentum expansion with $\Lmax=3$ and $\lonemax=\ltwomax=7$).
      Plot (a) shows that while there 
      is some noticeable change for short times, the results are stable at later times and seem to converge 
      to a limiting value. This is confirmed in the inset (b), which shows the same data vs.\@ $1/\tau$.
      Extrapolation to $1/\tau \to 0$ using a quadratic fit shows a limiting value of $0.4595$, only slightly higher 
      than the result obtained at $\tau=21\fs$.
      Plot (c) shows the temporal evolution of the ratio of the expectation values of the electron-electron interaction energy 
      $\expval{\op{H}_{12}}=\expval{\abs{\vecop{r}_1-\vecop{r}_2}^{-1}}$ and the total energy $\expval{\op{H}}$.}
  \label{fig:CS_conv_time}
\end{figure}

\hyperref[fig:CS_conv_time]{Figure \ref*{fig:CS_conv_time}} displays the convergence of the total cross section as a function of the field-free propagation time $\tau$. Delaying the projection from $\tau=1\fs$ to $\tau=21\fs$ changes the total cross section by less than $2\%$. Extrapolating to infinite time (and therefore to an infinite separation of the two electrons, 
\autoref{fig:CS_conv_time}b) changes the cross section by less than $0.2\%$. This gives an estimate of the error due to projection of that order of magnitude.
Furthermore, the electron-electron interaction energy is responsible for less than $1\%$ of the total energy of the wave packet at $\tau=21\fs$ (\autoref{fig:CS_conv_time}c).
The differential cross sections show the same qualitative convergence behavior with time as the total cross section.

This shows that projecting onto products of Coulomb waves is not a serious limitation. In other words, when the electrons have had time to move apart, their interaction can be neglected when projecting onto final states. For higher electron energies than in this test case ($\sim\!2.5\ev$),
the error is expected to be even smaller and the convergence faster.

\begin{figure*}[tbp]
  \includegraphics[width=0.485\linewidth]{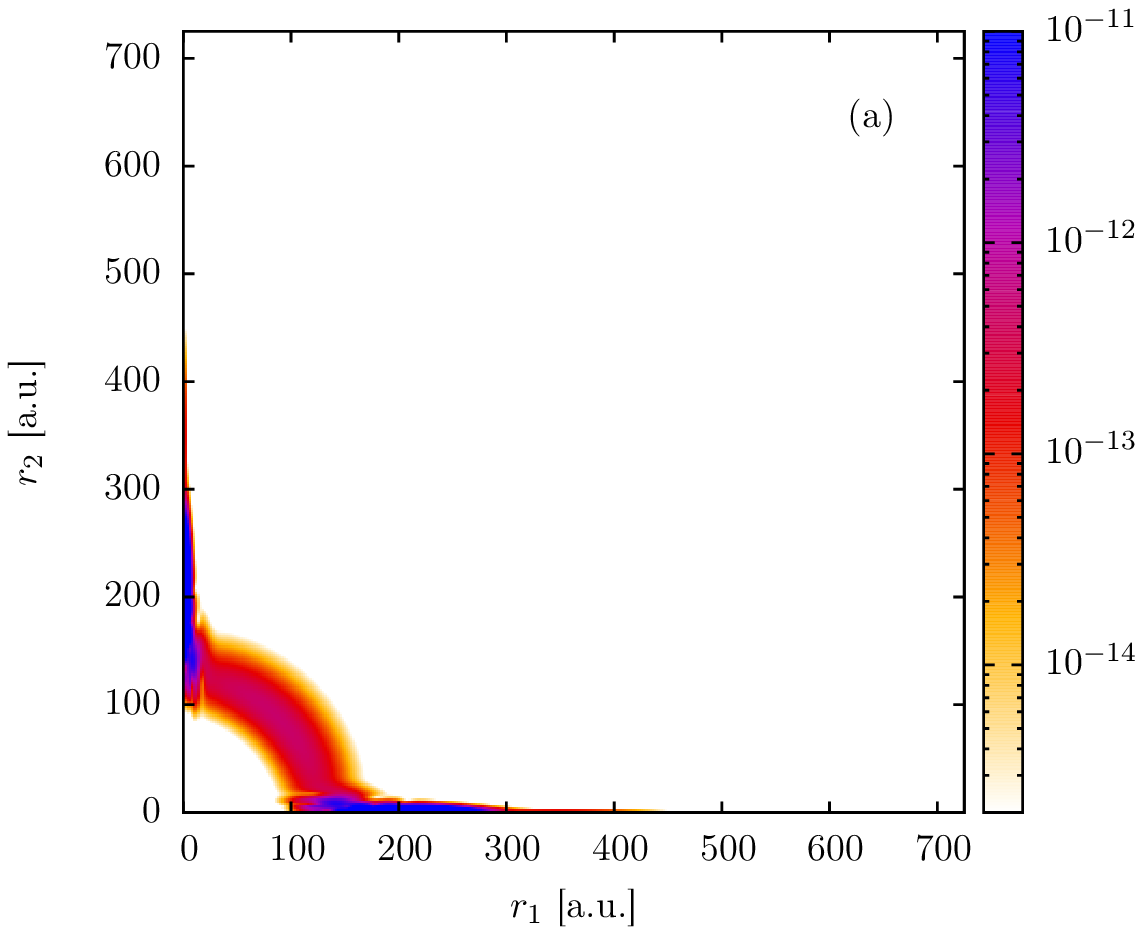}\hfill
  \includegraphics[width=0.485\linewidth]{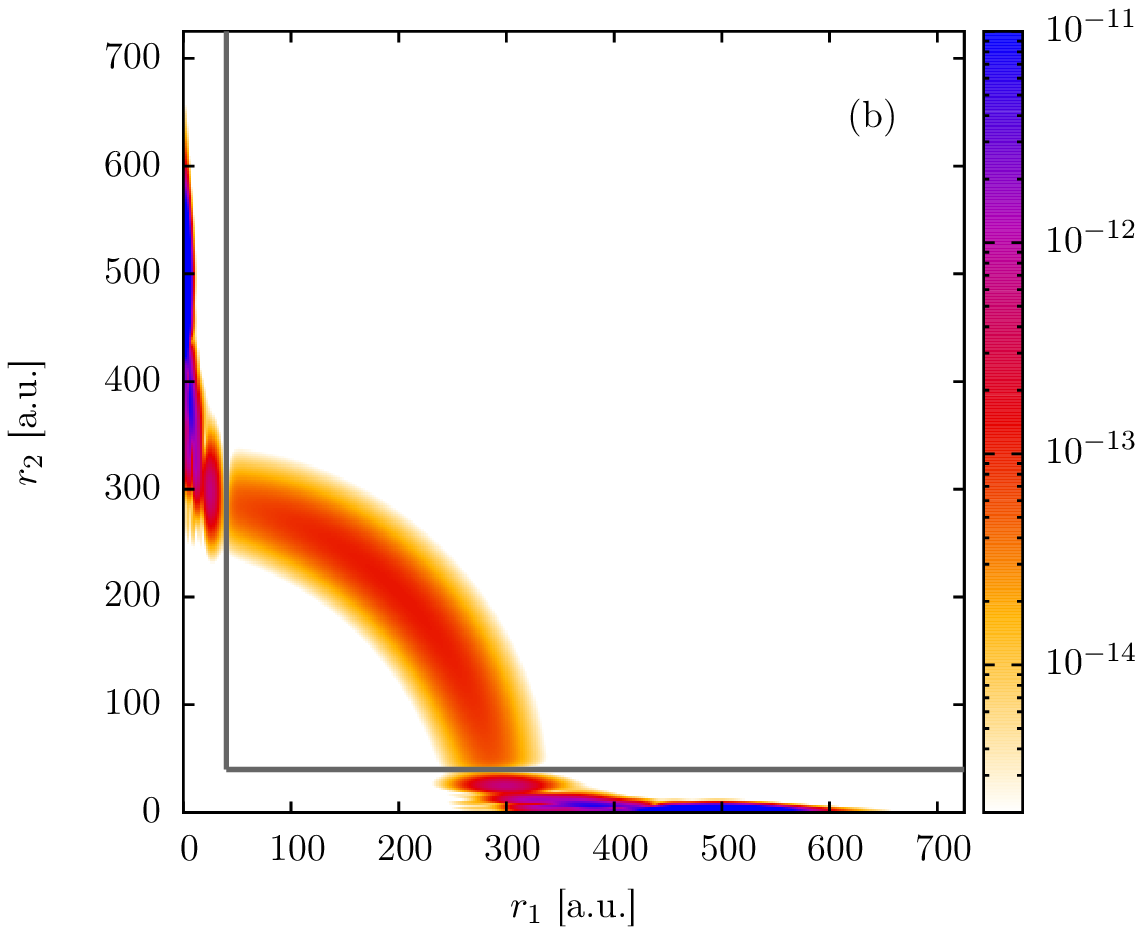}\\
  \includegraphics[width=0.485\linewidth]{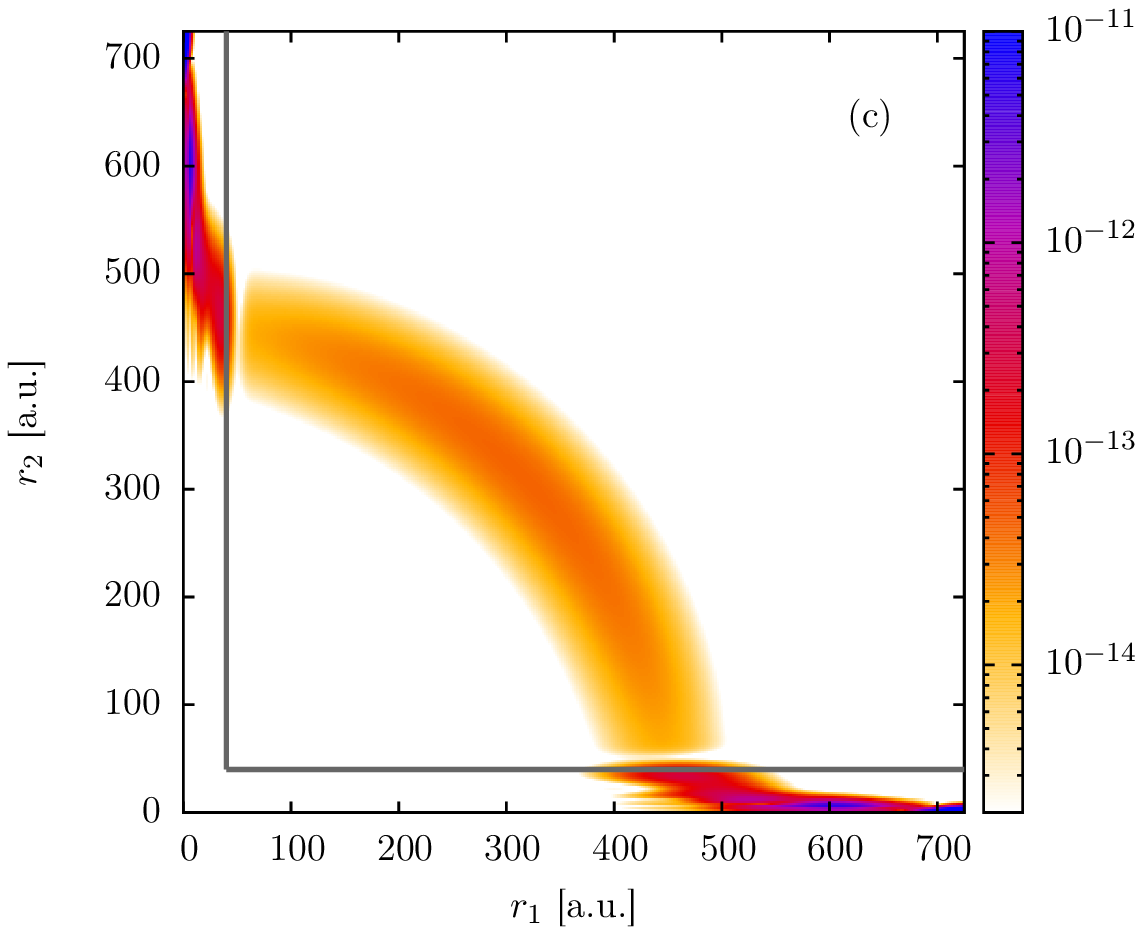}\hfill
  \includegraphics[width=0.485\linewidth]{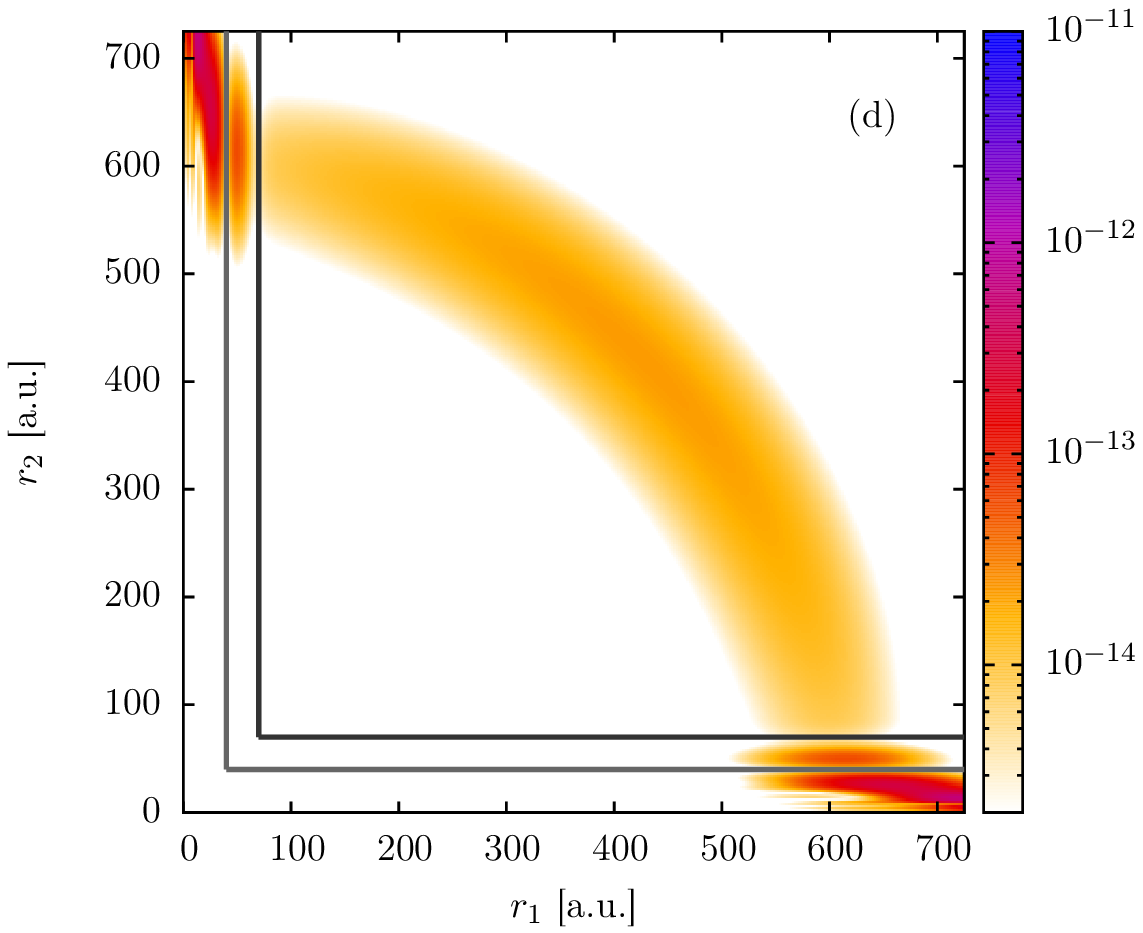}
  \caption{Radial wave function density for \autoref{fig:CS_conv_time}, at 
      $\tau=2\fs$ (a), $\tau=8\fs$ (b), $\tau=14\fs$ (c), and $\tau=20\fs$ (d) after the end of the pulse. For the wave function at $20\fs$,
      the doubly ionized part of the wave function is completely contained in the box with $r_1,r_2 > 70\au$\,. 
      By integrating the probability density over this region, an upper bound for the double 
      ionization yield can be established. The gray lines indicate the border between singly and 
      doubly ionized parts by visual inspection. The lighter gray line at $r_1,r_2=40\au$ is the apparent border at
      $\tau=8\fs$, while the darker gray line at $r_1,r_2=70\au$ is suggested by the distribution at $\tau=20\fs$. The density located between the two
      borders contains singly ionized parts that would by mistake be identified as being doubly ionized at $\tau=8\fs$ due
      to the lack of spatial separation.}
  \label{fig:visual_inspec_wf}
\end{figure*}

Due to the fact that the coordinate space representation of the fully correlated wave packet is available at the time of projection, an alternative, semi-quantitative check and error estimate for double ionization exists. From the visual inspection of snapshots of the joint radial distribution at different times (\autoref{fig:visual_inspec_wf}),
final states representing double ionization can be separated from those representing single ionization. While the singly ionized part of the wave function moves parallel to the $r_i$ axes, the doubly ionized parts of the wave function have positive momentum for both electrons so that they move away from both axes.
With increasing time, the spatial overlap between singly and doubly ionized states decreases, and the two contributions can be identified visually.
An estimate for an upper bound for the total double ionization cross section can thus be found by just integrating the radial density over the area that the doubly ionized wave packet occupies ($r_1,r_2 > 70\au$ in \autoref{fig:visual_inspec_wf}d).
This integral, which still contains a small portion of single ionization accompanied by excitation to Rydberg states, gives an \emph{upper bound} for the total double ionization cross section. In the present case (\autoref{fig:visual_inspec_wf}), the extracted estimate is about $25\%$ higher than the value determined by projection.
This discrepancy is predominantly caused by the existence of high-lying Rydberg states which also have contributions at large values of $r$.

\begin{figure*}[tbp]
  \includegraphics[width=0.48\linewidth]{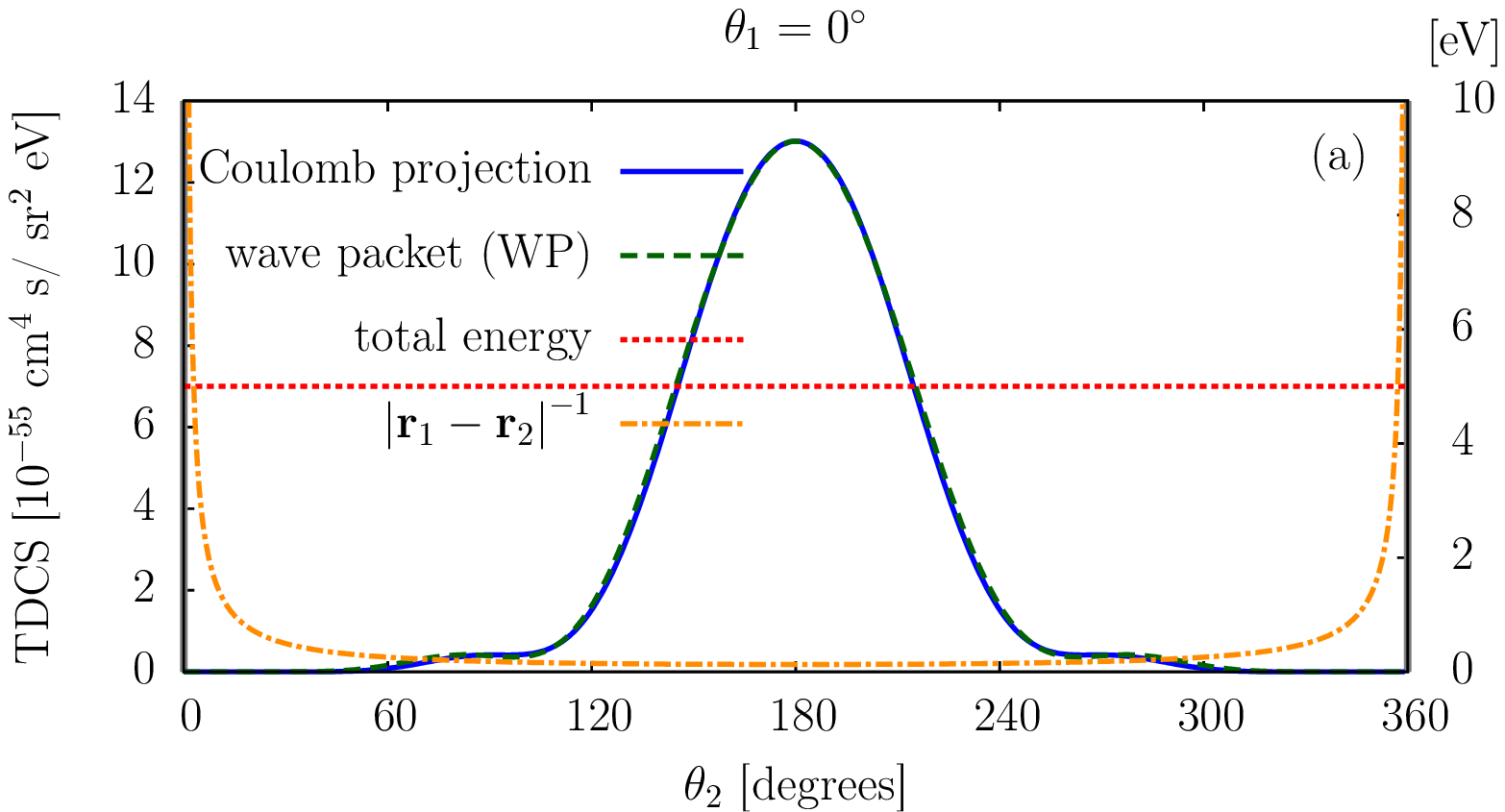}\hfill
  \includegraphics[width=0.48\linewidth]{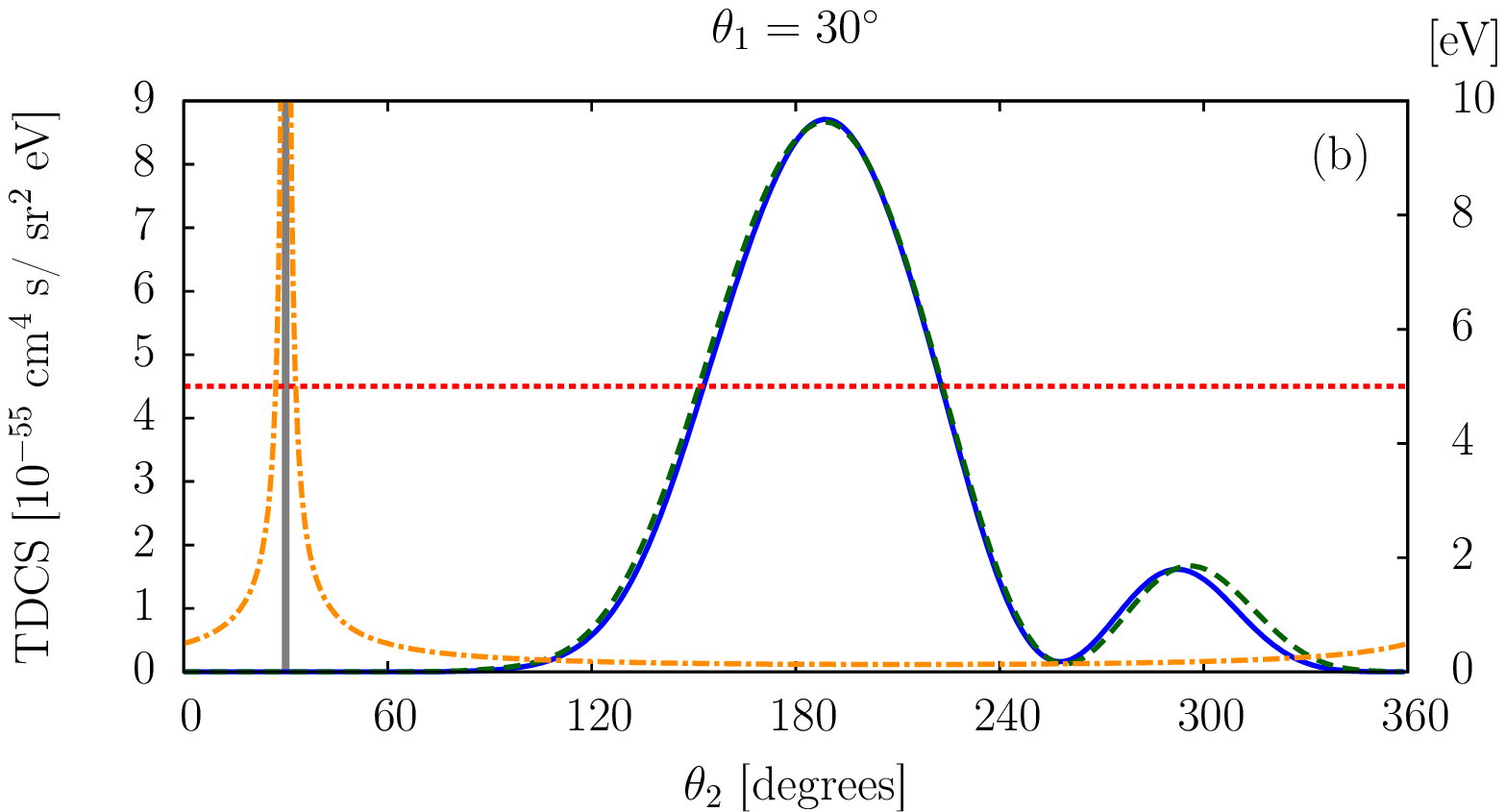}\\
  \vspace{2mm}
  \includegraphics[width=0.48\linewidth]{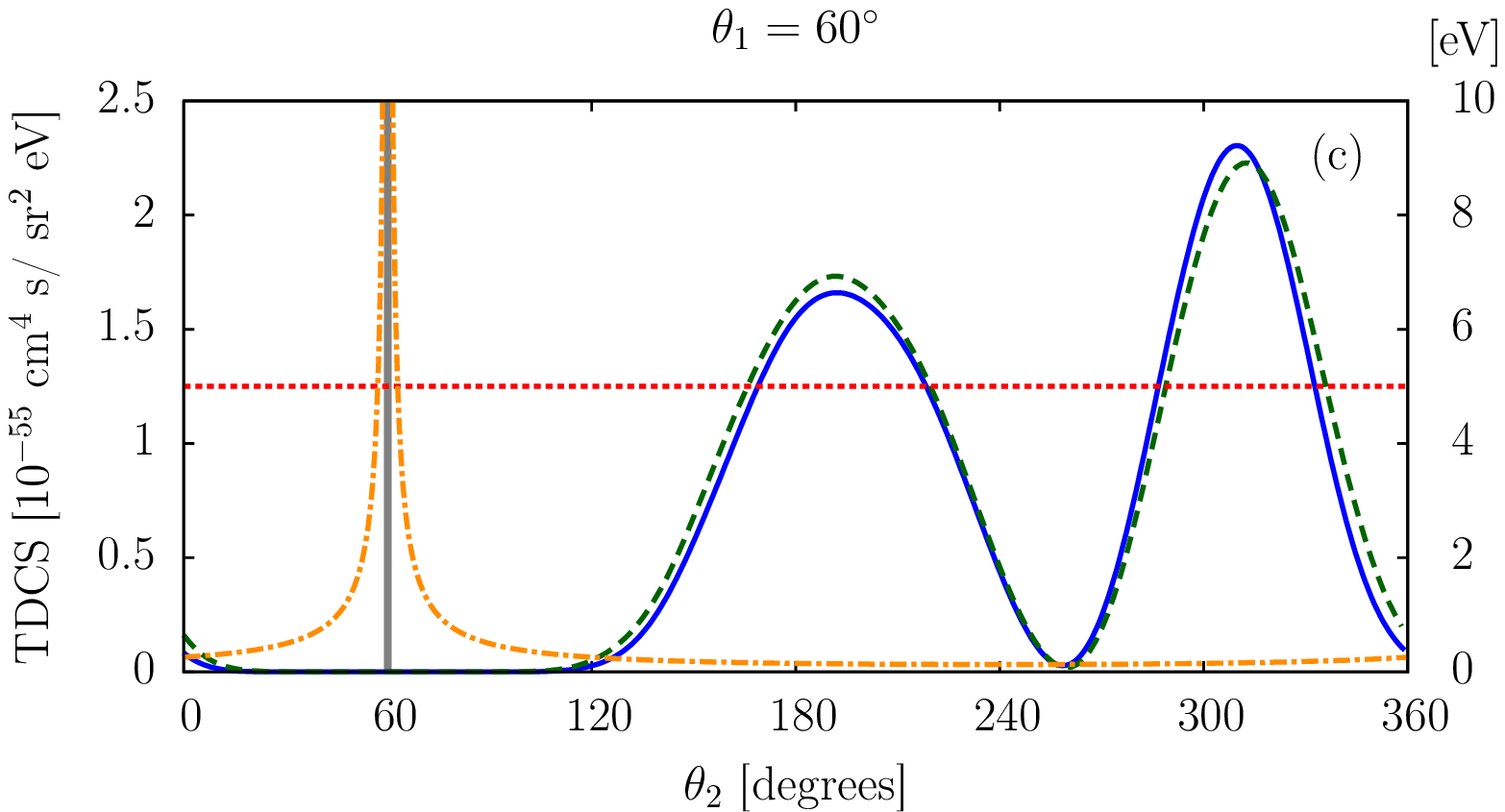}\hfill
  \includegraphics[width=0.48\linewidth]{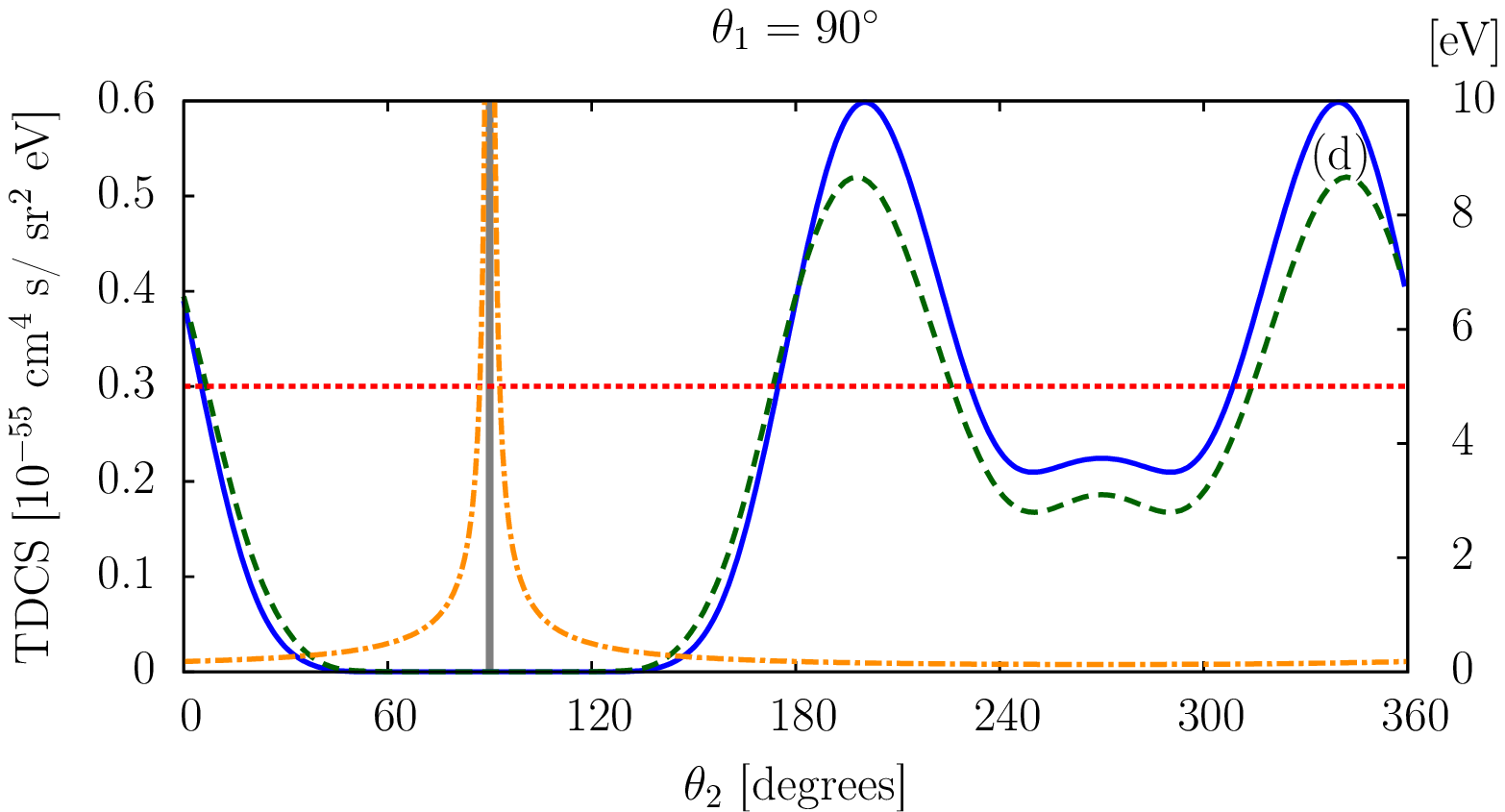}
  \caption{Comparison of different methods for extracting the triply differential cross 
      sections (TDCS) at $42\ev$ photon energy, with a $4\fs$ $\sin^2$ laser pulse.
      The data for Coulomb projection are obtained from \autoref{eq:tdcs_def} (at $E_1=2.5\ev$) 
      while the results labeled wave packet (WP) were obtained without transforming to 
      momentum space (\autoref{eq:tdcs_wp}). The angularly resolved value of 
      the electron-electron interaction energy at the position of the wave packet ($r_{1,2}\approx 150\au$) is also shown
      in comparison to the total energy of the doubly ionized wave packet ($\sim\!5\ev$).
      The vertical gray line shows the ejection angle $\theta_1$ of the first electron.
      The angular momentum expansion used values of $\Lmax=4$ and $\lonemax=\ltwomax=9$.
      The radial box had an extension of $400\au$, with FEDVR elements of $4\au$ and order $11$.}
  \label{fig:tdcs_wp}
\end{figure*}

One could expect that the choice of the final state is even more important when calculating differential cross 
sections because less degrees of freedom are integrated over. Specifically, the triply differential cross section (TDCS)
depends on the partial-wave phase shifts, which may not have fully converged at the time of projection.
In order to monitor possible errors in the angular distribution, we have also extracted the TDCS by a complementary 
method employing the coordinate representation of the two-electron wave packet at large propagation time, bypassing projection.
As we are interested in the TDCS at equal energy sharing, we take only that part of the wave packet with $r_1=r_2$, the 
part where both electrons have moved out to the same distance from the nucleus in the same time. This is what a (microscopic) 
time-of-flight detector would identify as equal-energy electrons. We then directly determine the angular distribution 
for this part of the wave function
\begin{equation}\label{eq:tdcs_wp}
\frac{\dd \sigma^\text{WP}(E_1=E_2)}{\dE_1 \dW_1 \dW_2} \propto \Int\dr\abs{\Psi(r,r,\Omega_1,\Omega_2)}^2 \,.
\end{equation}
This estimate for the TDCS, referred to in the following as the wave packet (WP) method, is compared with the projection
onto Coulomb waves (\autoref{eq:tdcs_def}) in \autoref{fig:tdcs_wp}. The excellent agreement we find attests to the fact 
that residual errors due to final-state correlations at the point of projection are, indeed, negligible.
Even when projecting onto plane waves (not shown), we have found that the TDCS at equal energy sharing almost exactly agrees with 
the result obtained from projection onto Coulomb waves (up to a global scaling factor of about $1.1$). This suggests 
that the Coulomb potential of the ionic core can also be neglected in the asymptotic region when only the 
differential behavior is of interest. The neglect of the electron-electron interaction is expected to be even less
important.

In addition, \autoref{fig:tdcs_wp} shows the angularly resolved value of the electron-electron interaction energy. 
For this, the radial distance of both electrons was taken as $r_{1,2}=150\au$, which corresponds to the position of
the doubly ionized wave packet at the point of projection. Clearly, the electrons only move into
directions where their interaction energy is negligible compared to the total energy of the doubly ionized wave packet.
This also supports the finding that the electron interaction can be neglected when doing the projection at late times.

The good agreement between the different methods used to extract total and differential cross sections can be understood 
by the following argument: The full wave function contains the electron correlation regardless of which basis it is expressed in.
The only ambiguity exists in identifying which parts of the wave packet at time $t=\tau$ will asymptotically correspond 
to the situation of interest (\ie double ionization in our case).
If $\tau$ is chosen large enough, the electrons have separated in space and their interaction energy 
is low (\cf \autoref{fig:CS_conv_time}c). This implies that the electrons will neither significantly deflect 
each other nor exchange energy at later times. Therefore, both the angular and the energy distribution are stable,
and the momenta of the electrons at time $t=\tau$ correspond to the asymptotic momenta for $t\to\infty$.
Similarly, channels where one of the electrons did not gain enough energy to escape the Coulomb potential 
of the nucleus by the time $t=\tau$ correspond to singly ionized final states, as the electron interaction
does not provide enough energy to change this situation at later times.

\subsection{Numerical convergence tests}\label{sec:conv}
In order to ensure the reliability of the calculated cross sections, we have performed extensive numerical testing and have found our results to be very well converged. The convergence issues addressed are (i) the radial discretization, (ii) the temporal propagation, and (iii) the angular momentum expansion. 

Convergence with respect to the radial grid (i) is easy to achieve within the FEDVR approach. All results shown were obtained with finite elements of $4\au$ extension and of order $11$. Results with order $13$ for elements of $4\au$ were virtually identical (within $0.02\%$ for the total cross sections). Convergence of the time propagation (ii) using our SIL method is equally uncritical. Even when relaxing the convergence criterion used for time propagation by two orders of magnitude, the results do not change perceptibly from those presented here. In addition, we also checked that our results do not depend on the gauge used in \autoref{eq:HeHam}. The change in the total cross section when switching from velocity gauge to length gauge is only $0.01\%$.
  
\begin{figure}[tbp]
  \includegraphics[width=0.99\linewidth]{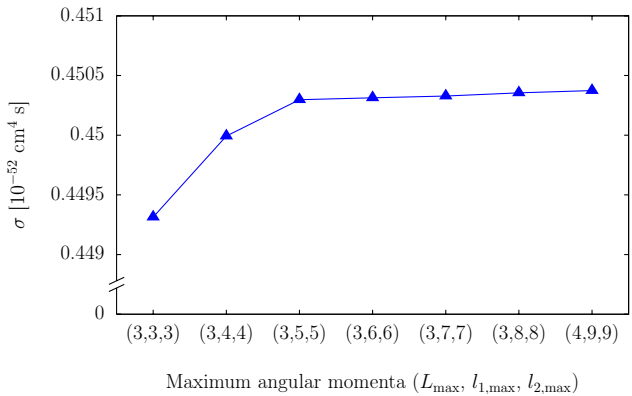}
  \caption{Convergence of the total cross section with size of angular momentum expansion. The total cross section is converged to within the accuracy
      of the method even for the smallest expansion in angular momenta. The differences in the result
      due to different angular basis sizes are much smaller than those observed when performing the projection at
      different times after the end of the pulse (\autoref{fig:CS_conv_time})
      ($4\fs$ $\sin^2$ pulse at $42\ev$ as in \autoref{fig:2dspectra})}.
  \label{fig:CS_conv_ang}
\end{figure}
  
The final question regarding convergence concerns the truncation of the angular momentum expansion \autoref{eq:pwe}. As the total angular momentum $L$ is conserved for the field-free Hamiltonian (because of spherical symmetry), the expansion does not require much higher values for $\Lmax$ than the minimum number of photons absorbed by the system. We have indeed found that there was no noticeable difference in any of the results between $\Lmax = 3$ or $\Lmax = 4$. Numerically, this is especially true when using velocity gauge. At the low intensities used here, the result is well converged with $\Lmax=3$ even when employing length gauge.
The convergence with respect to single particle angular momenta $(l_1,l_2)$, which are mixed by the electron-electron interaction, is much more critical.
The size of the expansion in $(l_1,l_2)$ strongly influences the accuracy of the angular distribution of the electrons and the degree of angular correlation.
      
\begin{figure*}[tbp]
  \includegraphics[width=0.48\linewidth]{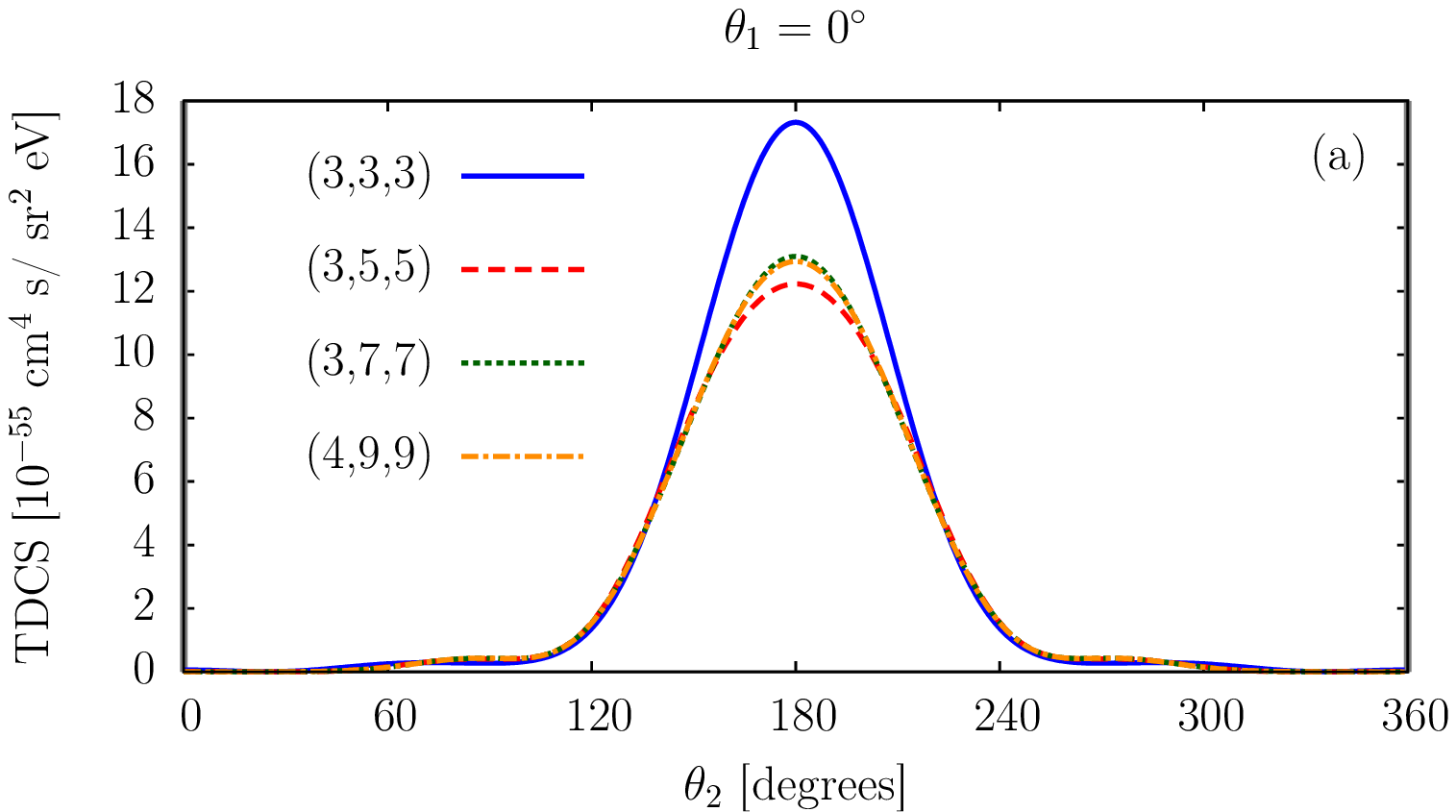}\hfill
  \includegraphics[width=0.48\linewidth]{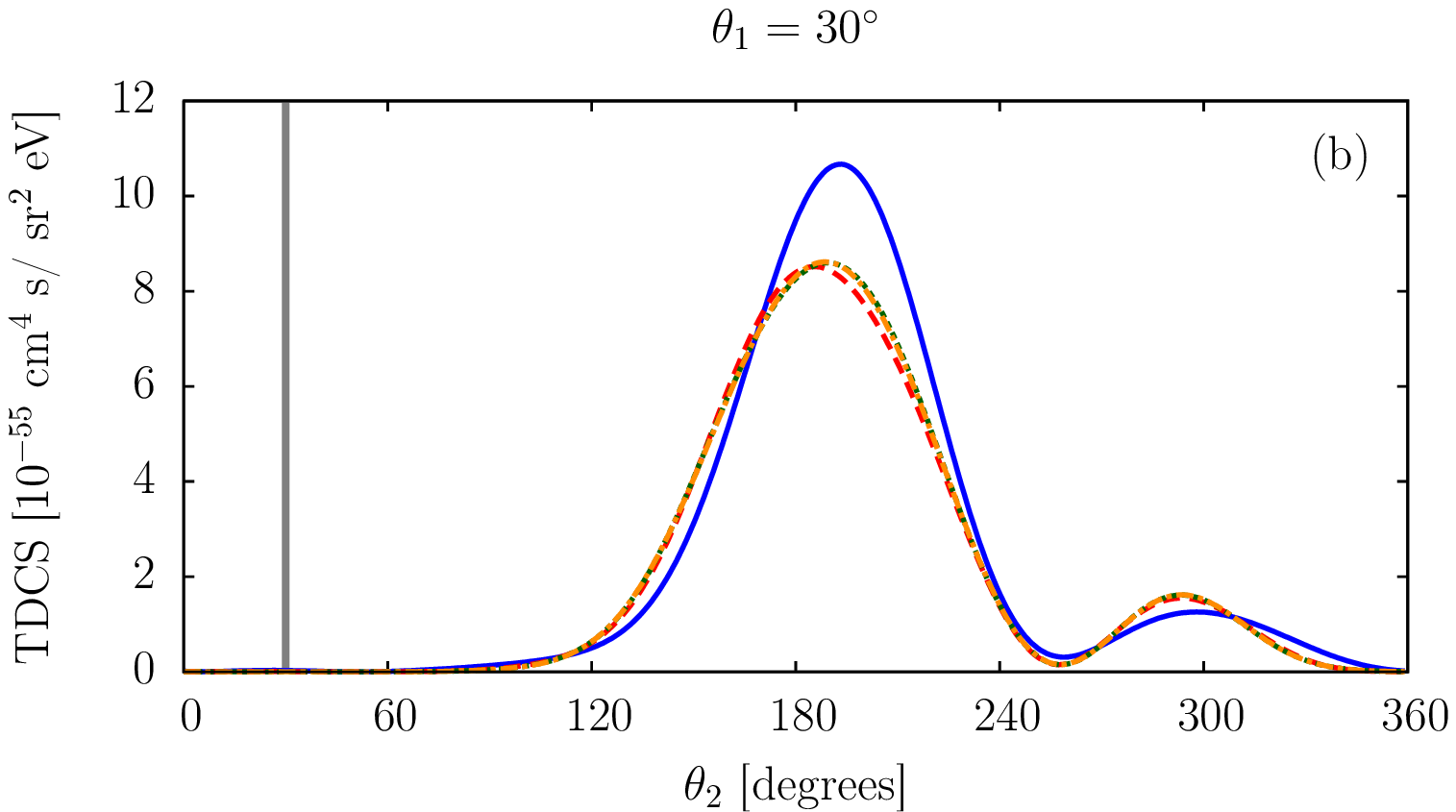}\\
  \vspace{2mm}
  \includegraphics[width=0.48\linewidth]{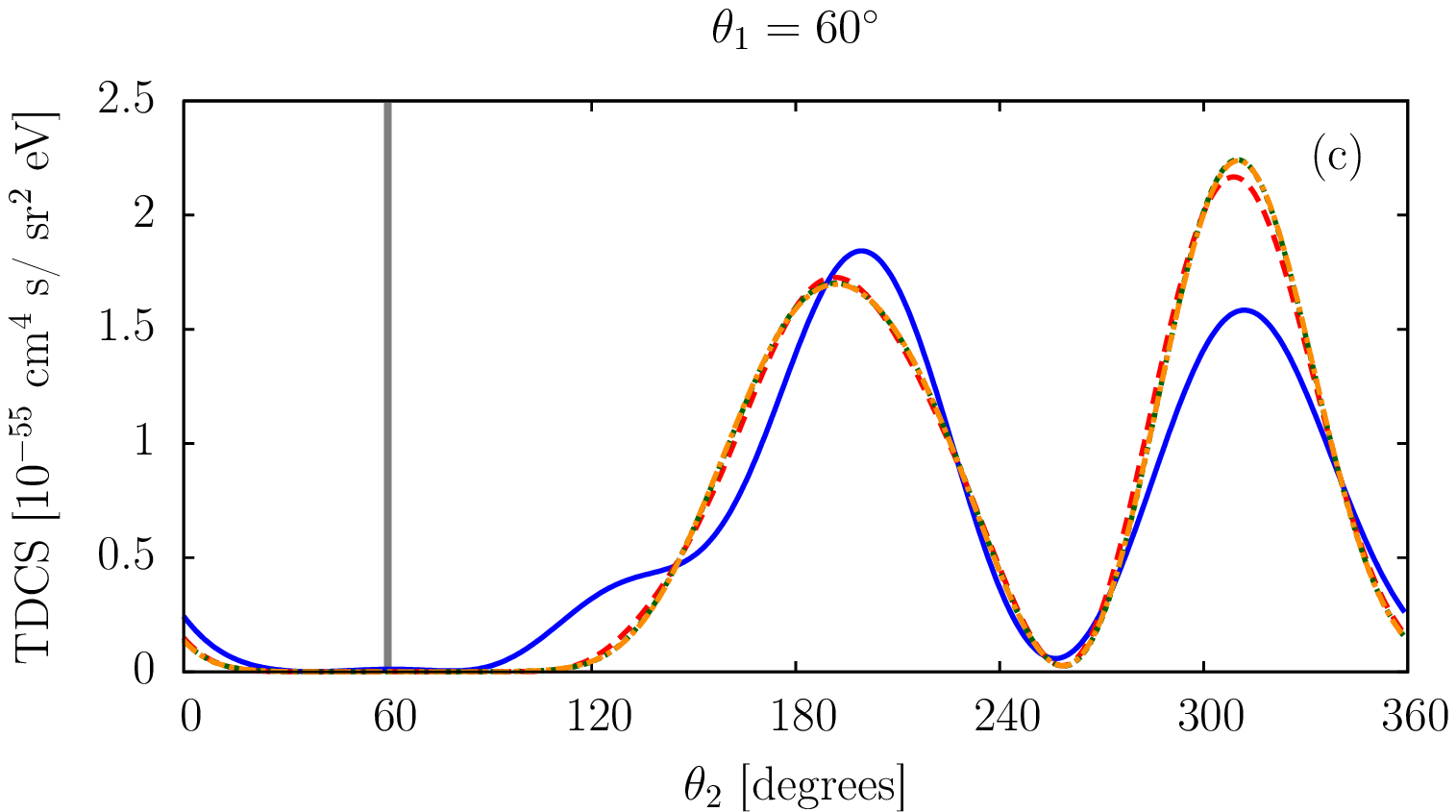}\hfill
  \includegraphics[width=0.48\linewidth]{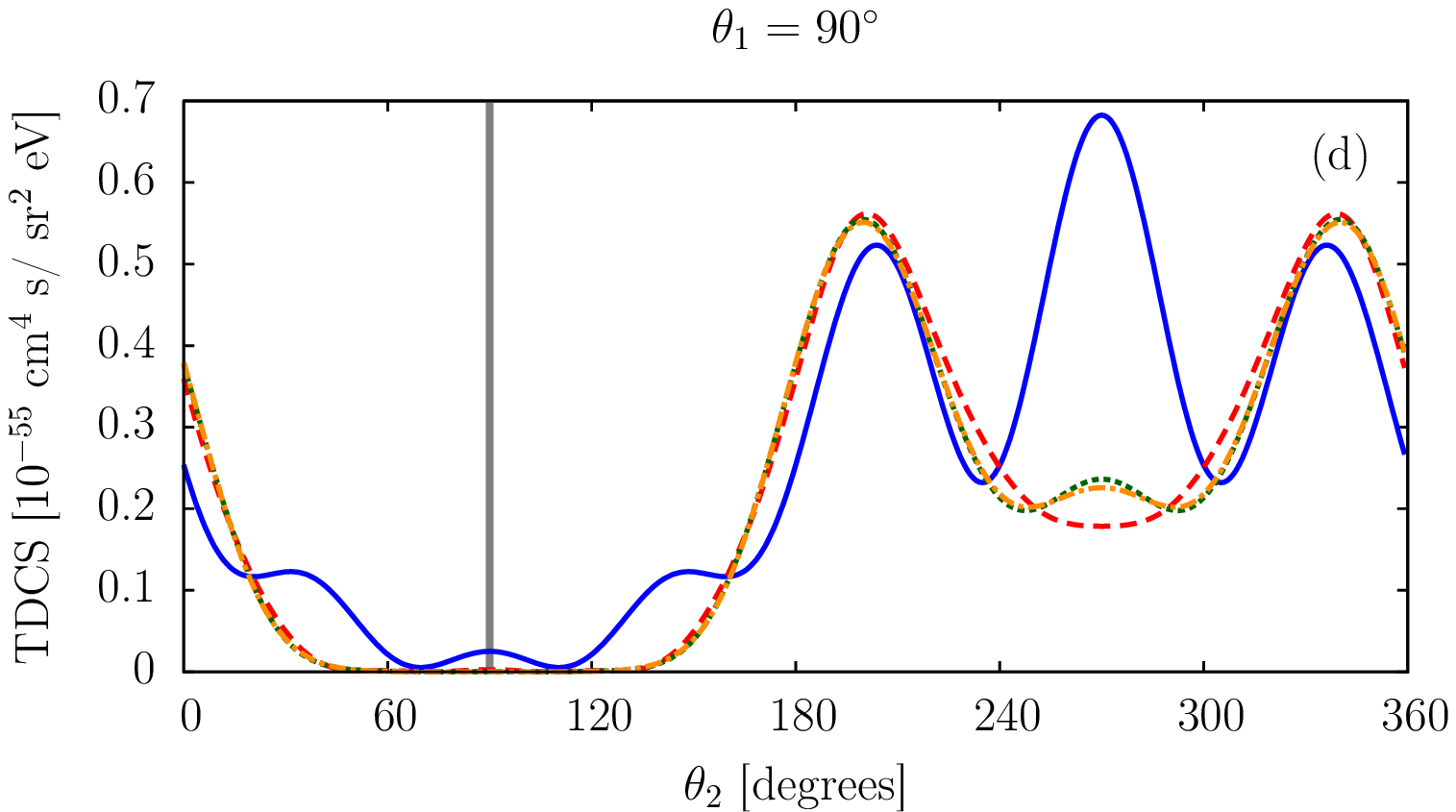}
  \caption{Convergence of the triply differential cross section (TDCS) with the size of the angular momentum expansion. The labels 
      specify the maximum values $(\Lmax,\lonemax,\ltwomax)$ used in the angular momentum expansion.
      The vertical gray line shows the ejection angle $\theta_1$ of the first electron.
      The TDCS converges only for relatively large values in the angular momentum expansion
      ($4\fs$ $\sin^2$ pulse at $42\ev$ as in \autoref{fig:2dspectra}).}
  \label{fig:tdcs_conv_ang}
\end{figure*}

While the total cross section, where all angles are integrated over, shows almost no dependence on the size of the angular momentum expansion, with variations of less than $0.3\%$ when $(\lonemax,\ltwomax)$ is increased from $(3,3)$ to $(9,9)$ (\autoref{fig:CS_conv_ang}), a different picture emerges when the two-electron angular distribution is considered.
The TDCS shows a strong dependence on the number of included partial waves.
For the present case, convergence is reached when single electron angular momenta up to $\lonemax=\ltwomax=7$ are included (see \autoref{fig:tdcs_conv_ang} below). Especially the TDCS at $\theta_1=90^\circ$ (where the cross section is very small) is very sensitive to the size of the partial wave expansion.

\section{Results}\label{sec:cs}\label{sec:CS}
\begin{figure*}[tbp]
  \includegraphics[width=0.48\linewidth]{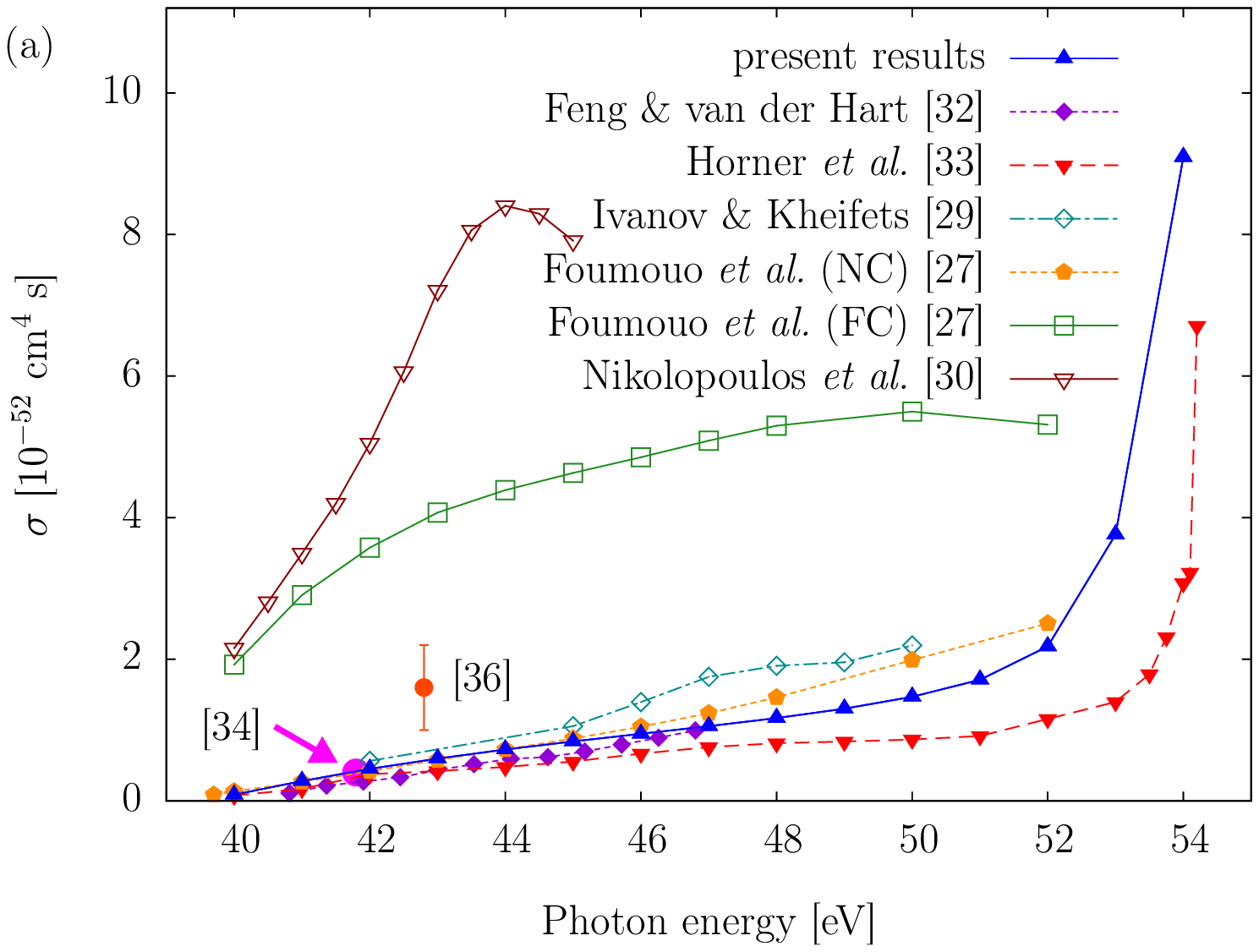}\hfill
  \includegraphics[width=0.48\linewidth]{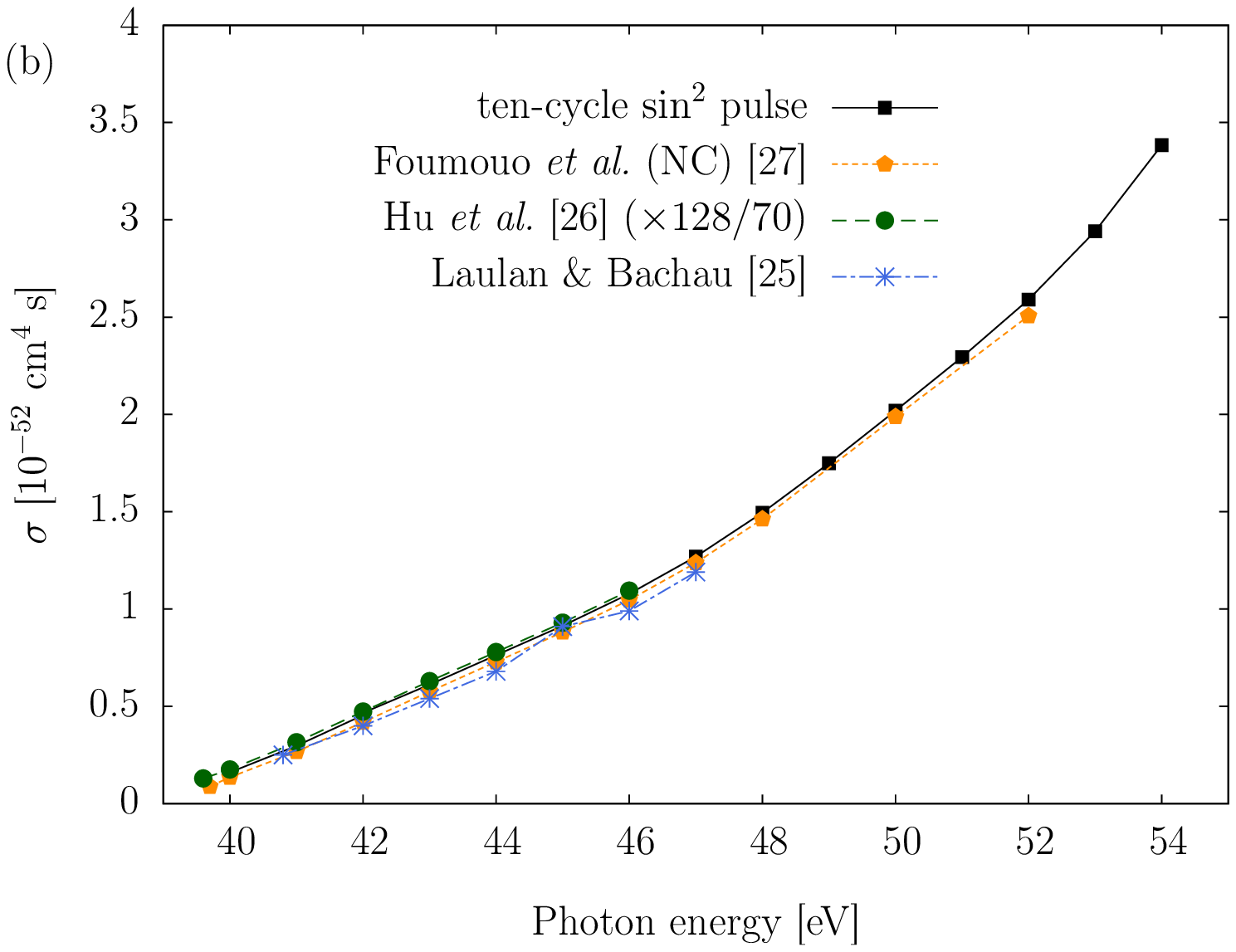}
  \caption{Comparison of the total two-photon double ionization (TPDI) cross sections, obtained from \autoref{eq:CS_def}, 
      with $\Teff=35T/128$.
      In (a), the laser pulses had a $\sin^2$ shape with a total duration of $4\fs$ and a peak
      intensity of $10^{12}\Wcm$. For the results of Foumouo \etal~\cite{FouLagEdaPir2006}, (NC) labels 
      the results obtained by projecting onto uncorrelated Coulomb waves, while (FC) labels the results 
      obtained using the $J$-matrix method. 
      (b) shows the comparison using ten-cycle ($\sim\!1\fs$) pulses to other time-dependent 
      approaches using the same pulses. Note that the results of Hu \etal~\cite{HuCoCo2005} were 
      rescaled by a factor of $128/70$ in order to include the correct $\Teff$.
      For both (a) and (b), the angular momenta were allowed to go up to $\Lmax=3$
      for the total angular momentum, and $\lonemax=\ltwomax=7$ for the single electron
      angular momenta. The radial box had an extension of $240\au$, with FEDVR elements
      of $4\au$ and order $11$.}
      \label{fig:CS}
\end{figure*}

In \autoref{fig:CS}, we compare the present results for the total cross section with various published data. The calculations were performed with a box size of $240\au$, with FEDVR elements that span $4\au$ and contain $11$ basis functions. The maximum angular momentum values are $\Lmax=3$ for the total angular momentum and $\lonemax=\ltwomax=7$ for the individual angular momenta. The laser pulse envelope had a $\sin^2$ shape, defined by 
\begin{equation}\label{eq:sin2_shape}
f(t) = \left\{ \begin{array}{ll} \sin^2\left(\frac{\pi}{T} t\right) & 0<t<T\\
                                 0                                & \text{otherwise} \end{array} \right. \,,
\end{equation}
with a total duration of $T=4\fs$ and a peak intensity of $I_0=10^{12}\Wcm$. The ionization yields were extracted $1\fs$ after the pulse. Following the results of \autoref{sec:final_corre} the projection error should not be larger than $2\%$. 

We compare our results with data from both time-dependent and time-independent approaches. Laulan and Bachau \cite{LauBac2003} solved the TDSE by means of a $B$-spline method and an explicit Runge-Kutta propagation scheme. The double ionization probability was obtained by projecting onto uncorrelated Coulomb functions. They also included first-order correction terms in the representation of the double continuum (thus partly taking into account radial correlations). However, they found little difference with respect to the uncorrelated functions. Hu, Colgan, and Collins \cite{HuCoCo2005} solved the time-dependent close-coupling equations using finite-difference techniques for the spatial discretization and the real-space product formula as well as a leapfrog algorithm for temporal propagation. The double ionization probability was also extracted by projection onto uncorrelated Coulomb waves. Foumouo \etal \cite{FouLagEdaPir2006} employed a spectral method of configuration interaction type (involving Coulomb-Sturmian functions) and an explicit Runge-Kutta time propagation to solve the TDSE. The double continuum was generated with the $J$-matrix method that should contain angular and radial correlations to the full extent. In addition, they also performed calculations using an uncorrelated representation of the two-electron continuum. The more recent results from Ivanov and Kheifets \cite{IvaKhe2007} are based on the time-dependent convergent close-coupling (CCC) method, taking into account correlations in the final state to some degree. Nikolopoulos and Lambropoulos \cite{NikLam2007} solved the TDSE using an expansion in correlated multichannel wave functions. 

Within the time-independent methods, Nikolopoulos and Lambropoulos \cite{NikLam2001} applied lowest-order non-vanishing perturbation theory (LOPT) to determine the generalized cross sections. Feng and van der Hart \cite{FenHar2003} employed $R$-matrix Floquet theory in combination with $B$-splines basis sets. The data from Horner \etal\cite{HorMorRes2007} also result from LOPT calculations. They solved the Dalgarno-Lewis equations for two-photon absorption in LOPT employing exterior complex scaling (ECS) and also account for correlation in initial, intermediate, and final states.

Overall, our results are in reasonable agreement with those of \cite{LauBac2003,HuCoCo2005,FouLagEdaPir2006,FenHar2003,HorMorRes2007} while sizable discrepancies exist in comparison with those of \cite{NikLam2001,NikLam2007} as well as those of \cite{FouLagEdaPir2006} in which corrections due to final-state correlations are included.
Clearly, the degree of convergence of the present results on the few percent level as well as the upper bound extracted from the radial wave packet analysis preclude any change of cross section by a factor of $5-10$, which would be necessary to obtain values of the same magnitude as \cite{NikLam2001,FouLagEdaPir2006,NikLam2007}.

The experimental values of Hasegawa, Nabekawa \etal\cite{HasTakNabIsh2005,NabHasTakMid2005} at $41.8\ev$ and of Sorokin \etal\cite{SorWelBob2007} at $42.8\ev$ (\cf \autoref{fig:CS}) are compatible with most of the theoretical data.
Due to the experimental uncertainties (\eg the harmonic intensity in \cite{HasTakNabIsh2005,NabHasTakMid2005} or the assumptions on the pulse shape and focusing conditions in \cite{SorWelBob2007}), the currently available data are not sufficient to strongly support or rule out any of the theoretical results.

The present results show a more pronounced variation with photon energy than other results obtained by direct integration of the time-dependent \Schro equation. This can be easily explained by the fact that most previous work employed ten-cycle pulses. At photon energies of $42$-$54\ev$, this corresponds to about $1\fs$ total duration, and consequently, a spectral width (FWHM) of about $6\ev$ (for $\sin^2$ pulses). The results are therefore an average over a rather large energy window. In contrast, we use pulses of $4\fs$ duration with a narrower spectrum (FWHM $\sim\!1.5\ev$).
To facilitate the comparison with previous calculations we have also performed a calculation for a ten-cycle pulse (\autoref{fig:CS}b) for which we find indeed better agreement.
The pulse duration dependence becomes, in particular, critical near the threshold for sequential ionization at $54.4\ev$. 

\begin{figure}
  \includegraphics[width=0.99\linewidth]{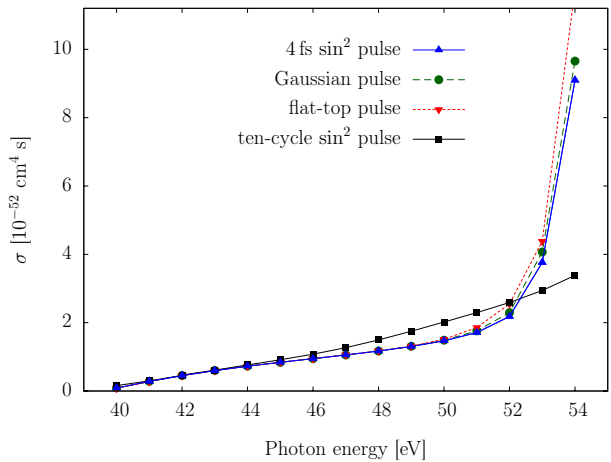}
  \caption{Total two-photon double ionization (TPDI) cross sections obtained using different pulse shapes.
      The duration of the Gaussian and flat-top pulse was chosen such that the FWHM of the spectral 
      distribution was identical to the one for the $4\fs$ $\sin^2$ pulse ($\sim\!1.5\ev$).
      The spectral FWHM of the ten-cycle ($\sim\!1\fs$) pulse is about four times larger, \ie $\sim\!6\ev$.
      All other parameters were the same as for \autoref{fig:CS}.}
  \label{fig:CS_pulses}
\end{figure}

For nonsequential processes, the yield is directly proportional to the duration of the pulse, so that the cross section can be defined as the proportionality factor between the double ionization rate and the photon flux $\Phi^N$, where $N$ is the number of photons for the direct process. On the other hand, the (two-photon) sequential ionization yield can be written as
\begin{equation}\label{eq:seq_yield}
P^{DI}_\text{seq} = \Int_{-\infty}^{\infty} \dt\, \sigma_{1} \Phi(t) \Int_{t}^{\infty} \dt' \sigma_{2} \Phi(t') \,,
\end{equation}
where $\sigma_{1}$ is the one-photon cross section for single ionization of He, and $\sigma_2$ is the one-photon cross section for ionization of the He$^+$ ion. Using the symmetry of the integrand yields
\begin{equation}\label{eq:seq_yield_2}
P^{DI}_\text{seq} = \sigma_1 \sigma_2 \frac{1}{2} \left(\Int_{-\infty}^{\infty} \dt \Phi(t) \right)^2 
             = \frac{\sigma_1 \sigma_2 I_0^2}{2\omega^2} (\Teff_{,1})^2 \,,
\end{equation}
which is proportional to the square of the total pulse duration $T$.
Proceeding along the same lines as for \autoref{eq:CS_def} by dividing the yield $P^{DI}_\text{seq}$ by the pulse duration results in an apparent ``cross section'' that increases linearly with the pulse length, contradicting the notion of a pulse shape and duration independent quantity. This is not surprising since for a two step process via on-shell intermediate states, a quadratic dependence on $\Teff$ (\autoref{eq:seq_yield_2}) is to be expected. If one extends the nonsequential cross section definition (\autoref{eq:CS_def}) into the threshold region for the sequential process, one expects a sudden rise whose height should be proportional to $\Teff$ and whose width is determined by the spectral broadening of the pulse.
With the pulses we used, the spectral width of $1.5\ev$ is small enough to observe the onset of this step discontinuity. In order to fully resolve the threshold behavior in a time-dependent calculation, even longer pulses with smaller bandwidth would be necessary. 

The region near the step discontinuity also provides a test case for the invariance of the nonsequential double ionization cross section under variation of the pulse shape. In addition to the $4\fs$ $\sin^2$ pulses used for most results shown in this paper, we also used the following pulse shapes: (i) a Gaussian pulse envelope and (ii) a flat-top pulse envelope with a $\sin^2$ ramp on for a quarter of the pulse duration, constant intensity for half the pulse duration, and a $\sin^2$ ramp off for the last quarter of the pulse. The durations of the Gaussian and flat-top pulses were chosen in such a way that the FWHM of the spectral distribution of the three pulse shapes was identical. Although all three pulses have the same spectral width, the distributions look different. Specifically, the spectral distribution of the flat-top pulse contains significant side lobes (ringing). In \autoref{fig:CS_pulses}, we show that the results obtained for the total cross section are almost identical with all three pulse shapes, apart from close to the step discontinuity at the threshold for sequential double ionization. Note that $\Teff$ is dependent on the pulse shape and has to be taken into account properly.

\begin{figure*}[tbp]
  \includegraphics[width=0.48\linewidth]{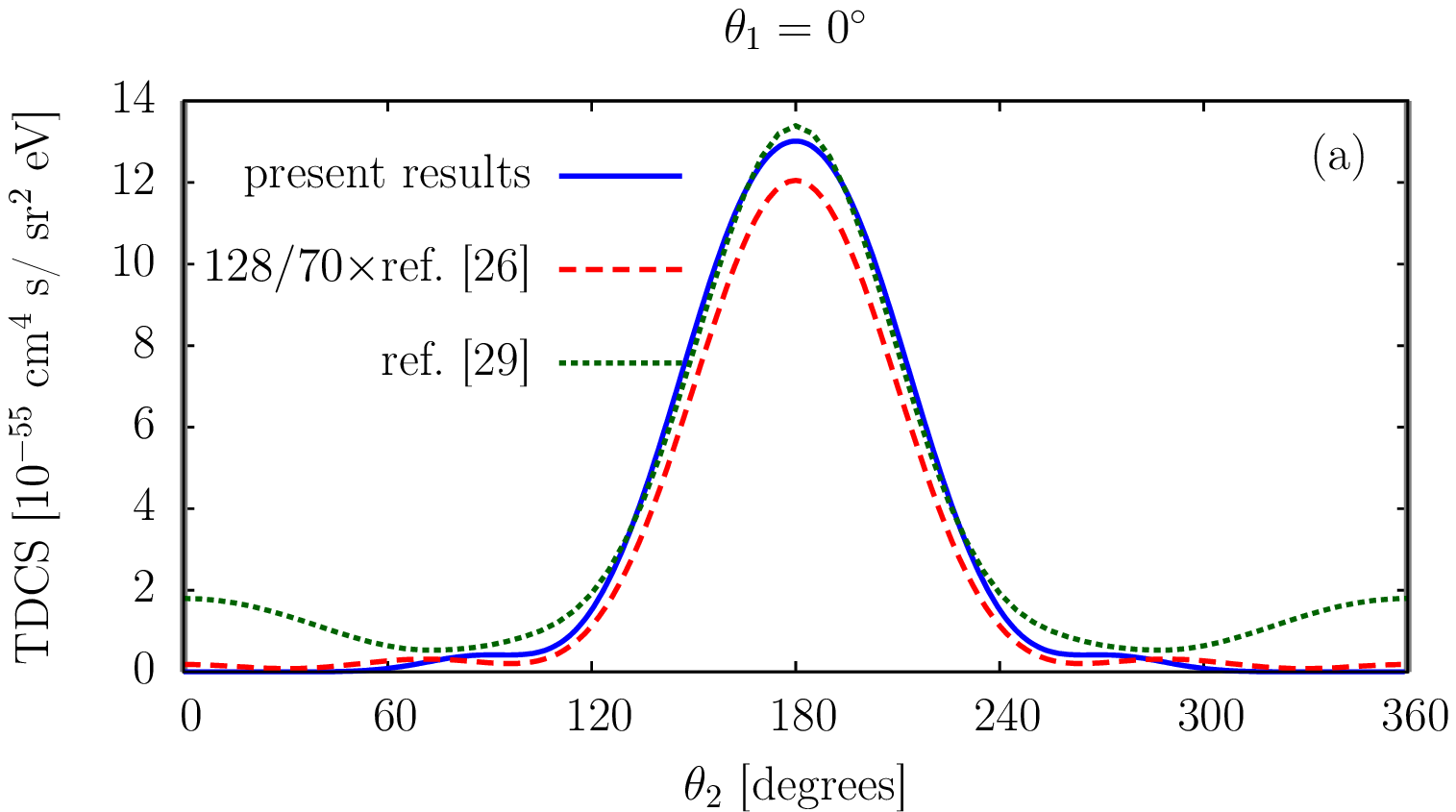}\hfill
  \includegraphics[width=0.48\linewidth]{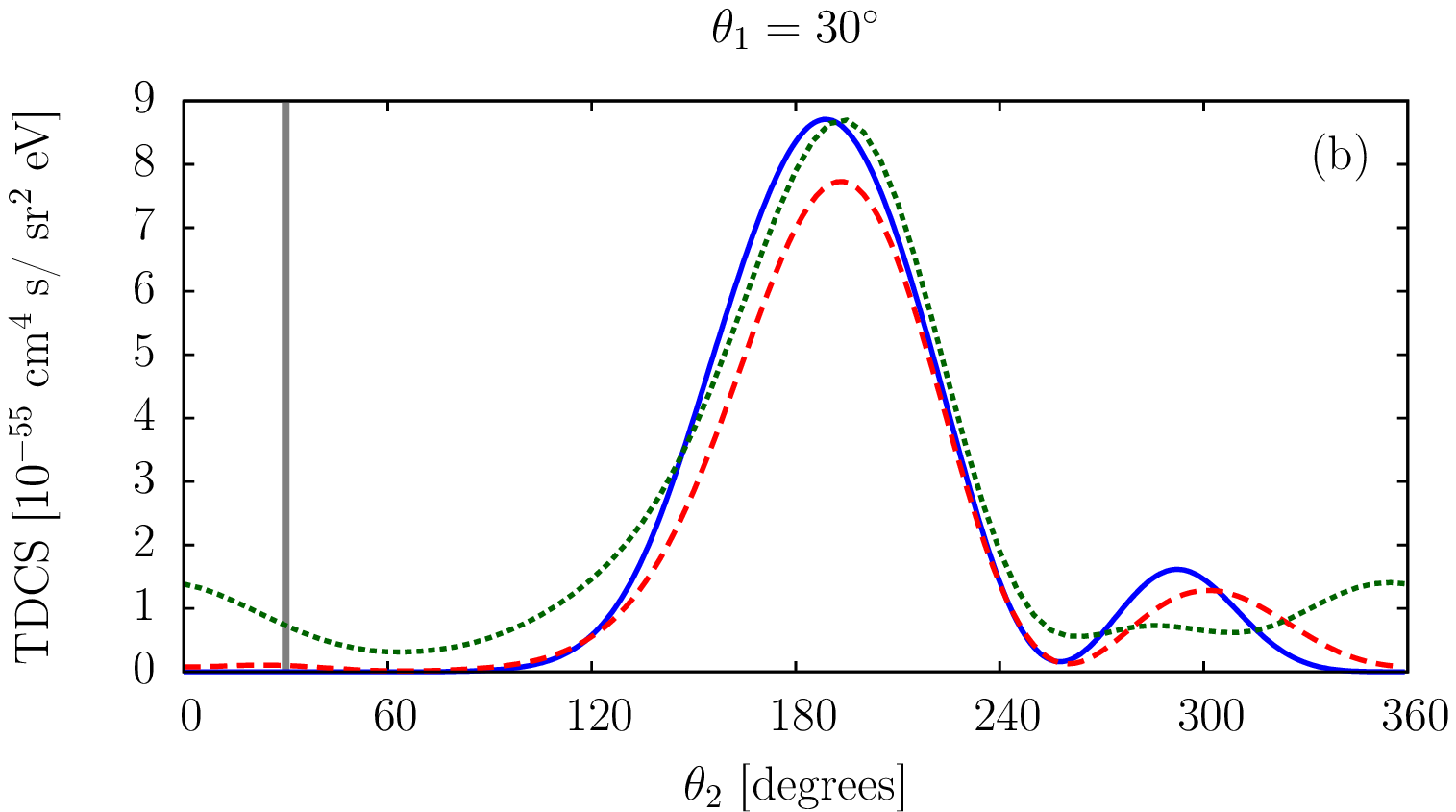}\\
  \vspace{2mm}
  \includegraphics[width=0.48\linewidth]{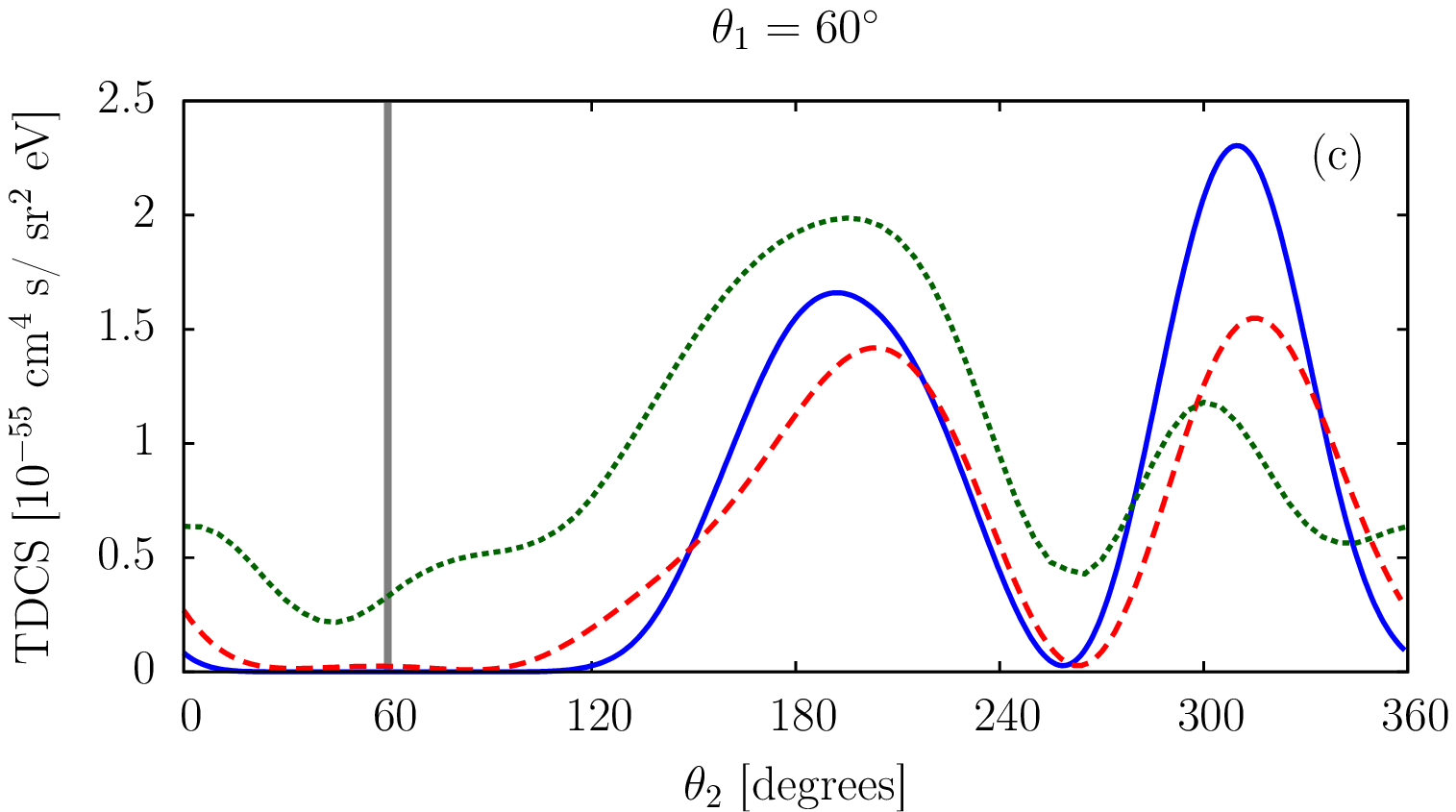}\hfill
  \includegraphics[width=0.48\linewidth]{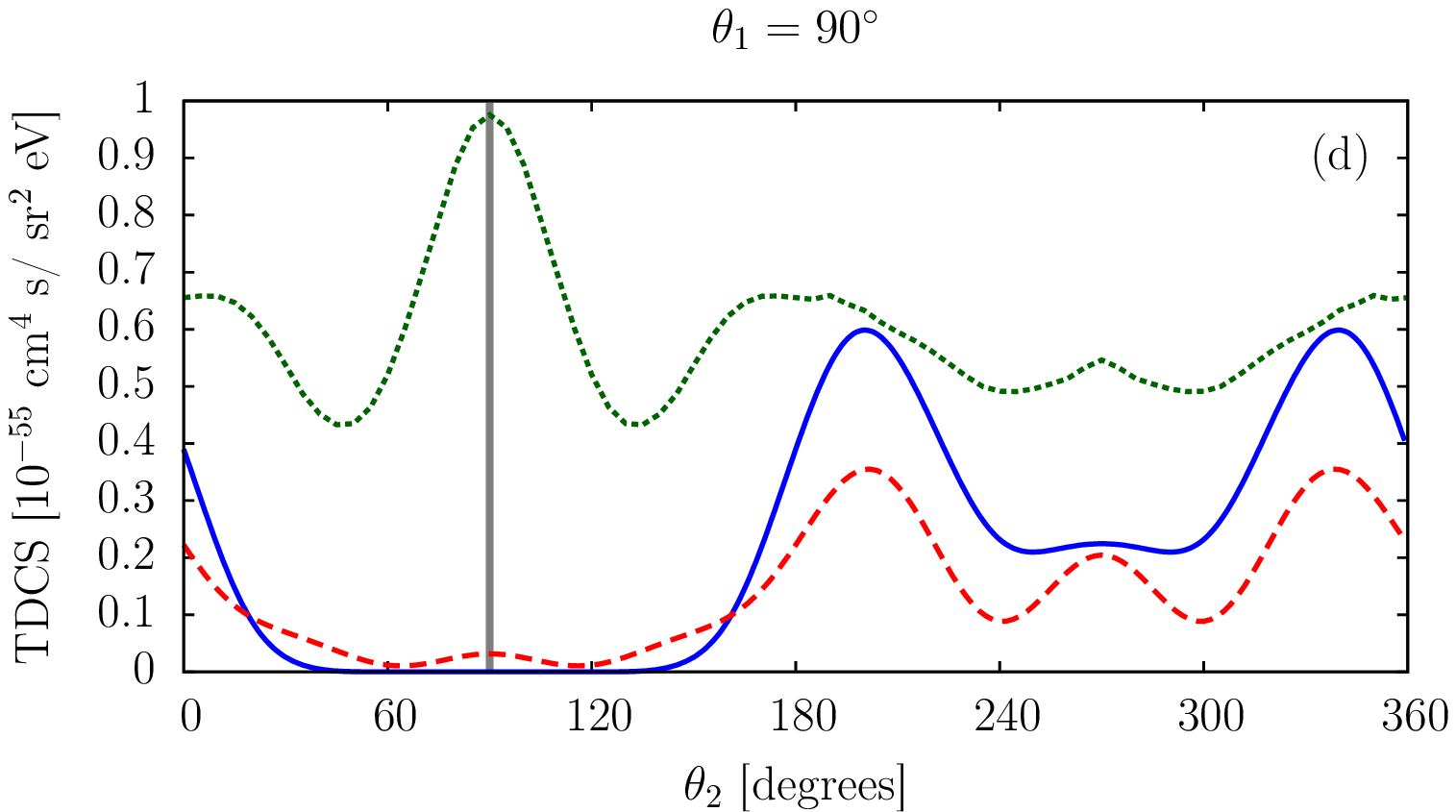}
  \caption{Comparison of triply differential cross sections (TDCS) at $42\ev$ photon energy. Our data are obtained 
      from \autoref{eq:tdcs_def} (at $E_1=2.5\ev$), with a $4\fs$ $\sin^2$ laser pulse.
      In comparison, the results of Hu \etal \cite{HuCoCo2005} and Ivanov and Kheifets \cite{IvaKhe2007} are shown.
      The vertical gray line shows the ejection angle $\theta_1$ of the first electron.
      The angular momentum expansion used values of $\Lmax=4$ and $\lonemax=\ltwomax=9$.
      The radial box had an extension of $400\au$, with FEDVR elements of $4\au$ and order $11$.}
  \label{fig:tdcs}
\end{figure*}

We turn now to the TDCS at $42\ev$, the quantity most sensitive to the level of the underlying approximations. The present results show qualitative agreement with some of the published data \cite{HuCoCo2005,IvaKhe2007}, but there are pronounced quantitative differences. While the prominent back-to-back emission lobe (anti-)parallel to the laser polarization direction is well reproduced in most calculations (\autoref{fig:tdcs}), the angular distribution for less favored emission directions (\eg $\theta_1=90^\circ$) differs significantly from other calculations. One reason is the sensitivity to the partial-wave expansion.
In contrast to the \emph{total} cross section, the TDCS needs a larger number of angular momentum combinations $(L,l_1,l_2)$ in the expansion of the wave function to converge. In order to resolve angular correlations, \ie for the triply differential cross section (TDCS), it is necessary to use relatively large expansions in single electron angular momenta. More specifically, good convergence of the TDCS is only reached for values as high as $\lonemax\!=\ltwomax\!=7$ (\cf \autoref{fig:tdcs_conv_ang}), which exceeds the angular momentum content of most other calculations.
As discussed in \autoref{sec:final_corre}, we have alternatively determined the TDCS by directly analyzing the angular distribution of the wave packet for equal energy sharing by a radial integral constrained to equal radii. We find remarkably close agreement with the Coulomb projection method (\autoref{fig:tdcs_wp}). The residual small deviations can be taken as an estimate for the uncertainty of the extraction method of the TDCS by Coulomb projection.

Ivanov and Kheifets~\cite{IvaKhe2007} take correlation in the final states into account using a convergent close-coupling
(CCC) method. While the magnitude of their results is similar to those presented here, the shape differs considerably.
In particular, they find significant probability for emission of both electrons in the same direction ($\theta_1=\theta_2$),
where the mutual repulsion of the electrons should be strongest.
In a very recent publication, Foumouo \etal\cite{FouAntPir2008} calculated the TDCS for equal energy sharing at 
$45\ev$ photon energy using two different methods. The results obtained by projecting the final wave function on products of 
Coulomb waves resemble ours (not shown here for $45\ev$, but the behavior is similar as for $42\ev$). 
However, when correlation in the final state is taken into account using a $J$-matrix method, the results are much larger in 
magnitude (as for the total cross section, \cf \autoref{fig:CS}) and display a shape reminiscent of the one obtained 
by Ivanov and Kheifets~\cite{IvaKhe2007}.

\section{Summary}
We have determined well-converged results for the total and triply differential (generalized) cross sections for nonsequential two-photon double ionization of helium. The total cross sections agree reasonably well with a number of recently published papers \cite{HuCoCo2005,HorMorRes2007,FenHar2003,IvaKhe2007}, but disagree with \cite{NikLam2001,NikLam2007}. While the uncorrelated results of \cite{FouLagEdaPir2006} fit well with our data, the $J$-matrix results that account for correlation, also presented in \cite{FouLagEdaPir2006}, are larger by almost an order of magnitude. In our approach, the inclusion of correlation in the final double continuum states is bypassed by waiting long enough after the end of the pulse before performing the projection onto uncorrelated final states. 

We have also presented approximate methods to extract both triply differential cross sections (TDCS) and total cross sections for nonsequential double ionization directly from the wave packet in coordinate space, thereby completely avoiding projection onto uncorrelated final states. We achieve excellent agreement between these complementary methods providing thereby a measure for the reliability and accuracy of the calculated cross sections.

Additionally, we have analyzed the pulse length dependence of the cross sections. In most of the previous time-dependent approaches ten-cycle pulses have been employed for the generalized cross sections extracted from the ionization yields. The resulting broad spectral width of the short pulse then influences the form of the cross section. This becomes evident from our calculations with considerably longer pulses. This is especially true at photon energies above $\sim 50\ev$ near the threshold for sequential ionization (at $54.4\ev$). Our own results for ten-cycle pulses, where these variations are smeared out, agree very well with the uncorrelated data from \cite{FouLagEdaPir2006} and the results from \cite{HuCoCo2005}. We are also aware of a recent approach by Guan \etal\cite{GuaBarSch2008}, who obtain similar values for ten-cycle pulses as well.

\subsection*{ACKNOWLEDGEMENTS}
We thank K.~Bartschat and X.~Guan for useful conversations. We also thank S.~Hu, I.~Ivanov, P.~Lambropoulos, B.~Piraux, 
and T.~Rescigno for sending their results in numerical form. 
J.F.\@ acknowledges support from the Electron and Atomic Physics Division at NIST in the initial phases of this research. 
J.F.\@, S.N.\@, R.P.\@, E.P.\@, and J.B.\@ acknowledge support by the FWF-Austria, Grant No.\@ SFB016.
Computational time provided under Institutional Computing at Los Alamos.
The Los Alamos National Laboratory is operated by Los Alamos National Security, LLC 
for the National Nuclear Security Administration of the U.S.\@ Department of Energy under Contract No.~DE-AC52-06NA25396.

\FloatBarrier


\begin{thebibliography}{49}
\expandafter\ifx\csname natexlab\endcsname\relax\def\natexlab#1{#1}\fi
\expandafter\ifx\csname bibnamefont\endcsname\relax
  \def\bibnamefont#1{#1}\fi
\expandafter\ifx\csname bibfnamefont\endcsname\relax
  \def\bibfnamefont#1{#1}\fi
\expandafter\ifx\csname citenamefont\endcsname\relax
  \def\citenamefont#1{#1}\fi
\expandafter\ifx\csname url\endcsname\relax
  \def\url#1{\texttt{#1}}\fi
\expandafter\ifx\csname urlprefix\endcsname\relax\def\urlprefix{URL }\fi
\providecommand{\bibinfo}[2]{#2}
\providecommand{\eprint}[2][]{\url{#2}}

\bibitem[{\citenamefont{Byron and Joachain}(1967)}]{Byron67}
\bibinfo{author}{\bibfnamefont{F.~W.} \bibnamefont{Byron}} \bibnamefont{and}
  \bibinfo{author}{\bibfnamefont{C.~J.} \bibnamefont{Joachain}},
  \bibinfo{journal}{Phys. Rev.} \textbf{\bibinfo{volume}{164}},
  \bibinfo{pages}{1} (\bibinfo{year}{1967}).

\bibitem[{\citenamefont{\r{A}berg}(1970)}]{Abe1970}
\bibinfo{author}{\bibfnamefont{T.}~\bibnamefont{\r{A}berg}},
  \bibinfo{journal}{Phys. Rev. A} \textbf{\bibinfo{volume}{2}},
  \bibinfo{pages}{1726} (\bibinfo{year}{1970}).

\bibitem[{\citenamefont{Dalgarno and Sadeghpour}(1992)}]{DalSad1992}
\bibinfo{author}{\bibfnamefont{A.}~\bibnamefont{Dalgarno}} \bibnamefont{and}
  \bibinfo{author}{\bibfnamefont{H.~R.} \bibnamefont{Sadeghpour}},
  \bibinfo{journal}{Phys. Rev. A} \textbf{\bibinfo{volume}{46}},
  \bibinfo{pages}{R3591} (\bibinfo{year}{1992}).

\bibitem[{\citenamefont{Andersson and Burgd\"{o}rfer}(1993)}]{Burg92}
\bibinfo{author}{\bibfnamefont{L.~R.} \bibnamefont{Andersson}}
  \bibnamefont{and}
  \bibinfo{author}{\bibfnamefont{J.}~\bibnamefont{Burgd\"{o}rfer}},
  \bibinfo{journal}{Phys. Rev. Lett.} \textbf{\bibinfo{volume}{71}},
  \bibinfo{pages}{50} (\bibinfo{year}{1993}).

\bibitem[{\citenamefont{Proulx and Shakeshaft}(1993)}]{Proulx93}
\bibinfo{author}{\bibfnamefont{D.}~\bibnamefont{Proulx}} \bibnamefont{and}
  \bibinfo{author}{\bibfnamefont{R.}~\bibnamefont{Shakeshaft}},
  \bibinfo{journal}{Phys. Rev. A} \textbf{\bibinfo{volume}{48}},
  \bibinfo{pages}{R875} (\bibinfo{year}{1993}).

\bibitem[{\citenamefont{Pont and Shakeshaft}(1995)}]{Pont95}
\bibinfo{author}{\bibfnamefont{M.}~\bibnamefont{Pont}} \bibnamefont{and}
  \bibinfo{author}{\bibfnamefont{R.}~\bibnamefont{Shakeshaft}},
  \bibinfo{journal}{Phys. Rev. A} \textbf{\bibinfo{volume}{51}},
  \bibinfo{pages}{R2676} (\bibinfo{year}{1995}).

\bibitem[{\citenamefont{Pindzola and Robicheaux}(1998)}]{PindRob98}
\bibinfo{author}{\bibfnamefont{M.~S.} \bibnamefont{Pindzola}} \bibnamefont{and}
  \bibinfo{author}{\bibfnamefont{F.}~\bibnamefont{Robicheaux}},
  \bibinfo{journal}{J. Phys. B} \textbf{\bibinfo{volume}{31}},
  \bibinfo{pages}{L823} (\bibinfo{year}{1998}).

\bibitem[{\citenamefont{Dundas et~al.}(1999)\citenamefont{Dundas, Taylor,
  Parker, and Smyth}}]{DunTayParSmy1999}
\bibinfo{author}{\bibfnamefont{D.}~\bibnamefont{Dundas}},
  \bibinfo{author}{\bibfnamefont{K.~T.} \bibnamefont{Taylor}},
  \bibinfo{author}{\bibfnamefont{J.~S.} \bibnamefont{Parker}},
  \bibnamefont{and} \bibinfo{author}{\bibfnamefont{E.~S.} \bibnamefont{Smyth}},
  \bibinfo{journal}{J. Phys. B} \textbf{\bibinfo{volume}{32}},
  \bibinfo{pages}{L231} (\bibinfo{year}{1999}).

\bibitem[{\citenamefont{Kornberg and Lambropoulos}(1999)}]{KorLam1999}
\bibinfo{author}{\bibfnamefont{M.~A.} \bibnamefont{Kornberg}} \bibnamefont{and}
  \bibinfo{author}{\bibfnamefont{P.}~\bibnamefont{Lambropoulos}},
  \bibinfo{journal}{J. Phys. B} \textbf{\bibinfo{volume}{32}},
  \bibinfo{pages}{L603} (\bibinfo{year}{1999}).

\bibitem[{\citenamefont{Becker and Faisal}(1999)}]{Becker99}
\bibinfo{author}{\bibfnamefont{A.}~\bibnamefont{Becker}} \bibnamefont{and}
  \bibinfo{author}{\bibfnamefont{F.~H.~M.} \bibnamefont{Faisal}},
  \bibinfo{journal}{Phys. Rev. A} \textbf{\bibinfo{volume}{59}},
  \bibinfo{pages}{R1742} (\bibinfo{year}{1999}).

\bibitem[{\citenamefont{Lein et~al.}(2000)\citenamefont{Lein, Gross, and
  Engel}}]{Lein2000}
\bibinfo{author}{\bibfnamefont{M.}~\bibnamefont{Lein}},
  \bibinfo{author}{\bibfnamefont{E.~K.~U.} \bibnamefont{Gross}},
  \bibnamefont{and} \bibinfo{author}{\bibfnamefont{V.}~\bibnamefont{Engel}},
  \bibinfo{journal}{Phys. Rev. Lett.} \textbf{\bibinfo{volume}{85}},
  \bibinfo{pages}{4707} (\bibinfo{year}{2000}).

\bibitem[{\citenamefont{Parker et~al.}(2001)\citenamefont{Parker, Moore,
  Meharg, Dundas, and Taylor}}]{Parker2000}
\bibinfo{author}{\bibfnamefont{J.~S.} \bibnamefont{Parker}},
  \bibinfo{author}{\bibfnamefont{L.~R.} \bibnamefont{Moore}},
  \bibinfo{author}{\bibfnamefont{K.~J.} \bibnamefont{Meharg}},
  \bibinfo{author}{\bibfnamefont{D.}~\bibnamefont{Dundas}}, \bibnamefont{and}
  \bibinfo{author}{\bibfnamefont{K.~T.} \bibnamefont{Taylor}},
  \bibinfo{journal}{J. Phys. B} \textbf{\bibinfo{volume}{34}},
  \bibinfo{pages}{L69} (\bibinfo{year}{2001}).

\bibitem[{\citenamefont{Mercouris et~al.}(2001)\citenamefont{Mercouris,
  Haritos, and Nicolaides}}]{MerHarNic2001}
\bibinfo{author}{\bibfnamefont{T.}~\bibnamefont{Mercouris}},
  \bibinfo{author}{\bibfnamefont{C.}~\bibnamefont{Haritos}}, \bibnamefont{and}
  \bibinfo{author}{\bibfnamefont{C.~A.} \bibnamefont{Nicolaides}},
  \bibinfo{journal}{J. Phys. B} \textbf{\bibinfo{volume}{34}},
  \bibinfo{pages}{3789} (\bibinfo{year}{2001}).

\bibitem[{\citenamefont{Ishikawa and Midorikawa}(2005)}]{IshMid2005}
\bibinfo{author}{\bibfnamefont{K.~L.} \bibnamefont{Ishikawa}} \bibnamefont{and}
  \bibinfo{author}{\bibfnamefont{K.}~\bibnamefont{Midorikawa}},
  \bibinfo{journal}{Phys. Rev. A} \textbf{\bibinfo{volume}{72}},
  \bibinfo{pages}{013407} (\bibinfo{year}{2005}).

\bibitem[{\citenamefont{Istomin et~al.}(2006)\citenamefont{Istomin, Pronin,
  Manakov, Marmo, and Starace}}]{IstProManMarSta2006}
\bibinfo{author}{\bibfnamefont{A.~Y.} \bibnamefont{Istomin}},
  \bibinfo{author}{\bibfnamefont{E.~A.} \bibnamefont{Pronin}},
  \bibinfo{author}{\bibfnamefont{N.~L.} \bibnamefont{Manakov}},
  \bibinfo{author}{\bibfnamefont{S.~I.} \bibnamefont{Marmo}}, \bibnamefont{and}
  \bibinfo{author}{\bibfnamefont{A.~F.} \bibnamefont{Starace}},
  \bibinfo{journal}{Phys. Rev. Lett.} \textbf{\bibinfo{volume}{97}},
  \bibinfo{pages}{123002} (\bibinfo{year}{2006}).

\bibitem[{\citenamefont{Makris et~al.}(2001)\citenamefont{Makris, Nikolopoulos,
  and Lambropoulos}}]{MakNikLam2001}
\bibinfo{author}{\bibfnamefont{M.~G.} \bibnamefont{Makris}},
  \bibinfo{author}{\bibfnamefont{L.~A.~A.} \bibnamefont{Nikolopoulos}},
  \bibnamefont{and}
  \bibinfo{author}{\bibfnamefont{P.}~\bibnamefont{Lambropoulos}},
  \bibinfo{journal}{Europhys. Lett.} \textbf{\bibinfo{volume}{54}},
  \bibinfo{pages}{722} (\bibinfo{year}{2001}).

\bibitem[{\citenamefont{Ackermann et~al.}(2007)\citenamefont{Ackermann, Asova,
  Ayvazyan, Azima, Baboi, Bahr, Balandin, Beutner, Brandt, Bolzmann
  et~al.}}]{FLASH2007}
\bibinfo{author}{\bibfnamefont{W.}~\bibnamefont{Ackermann}},
  \bibinfo{author}{\bibfnamefont{G.}~\bibnamefont{Asova}},
  \bibinfo{author}{\bibfnamefont{V.}~\bibnamefont{Ayvazyan}},
  \bibinfo{author}{\bibfnamefont{A.}~\bibnamefont{Azima}},
  \bibinfo{author}{\bibfnamefont{N.}~\bibnamefont{Baboi}},
  \bibinfo{author}{\bibfnamefont{J.}~\bibnamefont{Bahr}},
  \bibinfo{author}{\bibfnamefont{V.}~\bibnamefont{Balandin}},
  \bibinfo{author}{\bibfnamefont{B.}~\bibnamefont{Beutner}},
  \bibinfo{author}{\bibfnamefont{A.}~\bibnamefont{Brandt}},
  \bibinfo{author}{\bibfnamefont{A.}~\bibnamefont{Bolzmann}},
  \bibnamefont{et~al.}, \bibinfo{journal}{Nat. Photonics}
  \textbf{\bibinfo{volume}{1}}, \bibinfo{pages}{336} (\bibinfo{year}{2007}).

\bibitem[{\citenamefont{Dromey et~al.}(2006)\citenamefont{Dromey, Zepf, Gopal,
  Lancaster, Wei, Krushelnick, Tatarakis, Vakakis, Moustaizis, Kodama
  et~al.}}]{Dromey06}
\bibinfo{author}{\bibfnamefont{B.}~\bibnamefont{Dromey}},
  \bibinfo{author}{\bibfnamefont{M.}~\bibnamefont{Zepf}},
  \bibinfo{author}{\bibfnamefont{A.}~\bibnamefont{Gopal}},
  \bibinfo{author}{\bibfnamefont{K.}~\bibnamefont{Lancaster}},
  \bibinfo{author}{\bibfnamefont{M.~S.} \bibnamefont{Wei}},
  \bibinfo{author}{\bibfnamefont{K.}~\bibnamefont{Krushelnick}},
  \bibinfo{author}{\bibfnamefont{M.}~\bibnamefont{Tatarakis}},
  \bibinfo{author}{\bibfnamefont{N.}~\bibnamefont{Vakakis}},
  \bibinfo{author}{\bibfnamefont{S.}~\bibnamefont{Moustaizis}},
  \bibinfo{author}{\bibfnamefont{R.}~\bibnamefont{Kodama}},
  \bibnamefont{et~al.}, \bibinfo{journal}{Nat. Phys.}
  \textbf{\bibinfo{volume}{2}}, \bibinfo{pages}{456} (\bibinfo{year}{2006}).

\bibitem[{\citenamefont{Naumova et~al.}(2004)\citenamefont{Naumova, Nees,
  Sokolov, Hou, and Mourou}}]{NauNeeSok2004}
\bibinfo{author}{\bibfnamefont{N.~M.} \bibnamefont{Naumova}},
  \bibinfo{author}{\bibfnamefont{J.~A.} \bibnamefont{Nees}},
  \bibinfo{author}{\bibfnamefont{I.~V.} \bibnamefont{Sokolov}},
  \bibinfo{author}{\bibfnamefont{B.}~\bibnamefont{Hou}}, \bibnamefont{and}
  \bibinfo{author}{\bibfnamefont{G.~A.} \bibnamefont{Mourou}},
  \bibinfo{journal}{Phys. Rev. Lett.} \textbf{\bibinfo{volume}{92}},
  \bibinfo{pages}{063902} (\bibinfo{year}{2004}).

\bibitem[{\citenamefont{Tsakiris et~al.}(2006)\citenamefont{Tsakiris, Eidmann,
  Meyer-Ter-Vehn, and Krausz}}]{Tsak06}
\bibinfo{author}{\bibfnamefont{G.~D.} \bibnamefont{Tsakiris}},
  \bibinfo{author}{\bibfnamefont{K.}~\bibnamefont{Eidmann}},
  \bibinfo{author}{\bibfnamefont{J.}~\bibnamefont{Meyer-Ter-Vehn}},
  \bibnamefont{and} \bibinfo{author}{\bibfnamefont{F.}~\bibnamefont{Krausz}},
  \bibinfo{journal}{New J. Phys.} \textbf{\bibinfo{volume}{8}},
  \bibinfo{pages}{19} (\bibinfo{year}{2006}).

\bibitem[{\citenamefont{Seres et~al.}(2007)\citenamefont{Seres, Yakovlev,
  Seres, Streli, Wobrauschek, Spielmann, and Krausz}}]{SerYakSer2007}
\bibinfo{author}{\bibfnamefont{J.}~\bibnamefont{Seres}},
  \bibinfo{author}{\bibfnamefont{V.~S.} \bibnamefont{Yakovlev}},
  \bibinfo{author}{\bibfnamefont{E.}~\bibnamefont{Seres}},
  \bibinfo{author}{\bibfnamefont{C.}~\bibnamefont{Streli}},
  \bibinfo{author}{\bibfnamefont{P.}~\bibnamefont{Wobrauschek}},
  \bibinfo{author}{\bibfnamefont{C.}~\bibnamefont{Spielmann}},
  \bibnamefont{and} \bibinfo{author}{\bibfnamefont{F.}~\bibnamefont{Krausz}},
  \bibinfo{journal}{Nat. Phys.} \textbf{\bibinfo{volume}{3}},
  \bibinfo{pages}{878} (\bibinfo{year}{2007}).

\bibitem[{\citenamefont{Gibson et~al.}(2003)\citenamefont{Gibson, Paul, Wagner,
  Tobey, Gaudiosi, Backus, Christov, Aquila, Gullikson, Attwood
  et~al.}}]{GibPauWag2003}
\bibinfo{author}{\bibfnamefont{E.~A.} \bibnamefont{Gibson}},
  \bibinfo{author}{\bibfnamefont{A.}~\bibnamefont{Paul}},
  \bibinfo{author}{\bibfnamefont{N.}~\bibnamefont{Wagner}},
  \bibinfo{author}{\bibfnamefont{R.}~\bibnamefont{Tobey}},
  \bibinfo{author}{\bibfnamefont{D.}~\bibnamefont{Gaudiosi}},
  \bibinfo{author}{\bibfnamefont{S.}~\bibnamefont{Backus}},
  \bibinfo{author}{\bibfnamefont{I.~P.} \bibnamefont{Christov}},
  \bibinfo{author}{\bibfnamefont{A.}~\bibnamefont{Aquila}},
  \bibinfo{author}{\bibfnamefont{E.~M.} \bibnamefont{Gullikson}},
  \bibinfo{author}{\bibfnamefont{D.~T.} \bibnamefont{Attwood}},
  \bibnamefont{et~al.}, \bibinfo{journal}{Science}
  \textbf{\bibinfo{volume}{302}}, \bibinfo{pages}{95} (\bibinfo{year}{2003}).

\bibitem[{\citenamefont{Zhang et~al.}(2007)\citenamefont{Zhang, Lytle,
  Popmintchev, Zhou, Kapteyn, Murnane, and Cohen}}]{ZhaLytPop2007}
\bibinfo{author}{\bibfnamefont{X.}~\bibnamefont{Zhang}},
  \bibinfo{author}{\bibfnamefont{A.~L.} \bibnamefont{Lytle}},
  \bibinfo{author}{\bibfnamefont{T.}~\bibnamefont{Popmintchev}},
  \bibinfo{author}{\bibfnamefont{X.}~\bibnamefont{Zhou}},
  \bibinfo{author}{\bibfnamefont{H.~C.} \bibnamefont{Kapteyn}},
  \bibinfo{author}{\bibfnamefont{M.~M.} \bibnamefont{Murnane}},
  \bibnamefont{and} \bibinfo{author}{\bibfnamefont{O.}~\bibnamefont{Cohen}},
  \bibinfo{journal}{Nat. Phys.} \textbf{\bibinfo{volume}{3}},
  \bibinfo{pages}{270} (\bibinfo{year}{2007}).

\bibitem[{\citenamefont{Colgan and Pindzola}(2002)}]{ColPin2002}
\bibinfo{author}{\bibfnamefont{J.}~\bibnamefont{Colgan}} \bibnamefont{and}
  \bibinfo{author}{\bibfnamefont{M.~S.} \bibnamefont{Pindzola}},
  \bibinfo{journal}{Phys. Rev. Lett.} \textbf{\bibinfo{volume}{88}},
  \bibinfo{pages}{173002} (\bibinfo{year}{2002}).

\bibitem[{\citenamefont{Laulan and Bachau}(2003)}]{LauBac2003}
\bibinfo{author}{\bibfnamefont{S.}~\bibnamefont{Laulan}} \bibnamefont{and}
  \bibinfo{author}{\bibfnamefont{H.}~\bibnamefont{Bachau}},
  \bibinfo{journal}{Phys. Rev. A} \textbf{\bibinfo{volume}{68}},
  \bibinfo{pages}{013409} (\bibinfo{year}{2003}).

\bibitem[{\citenamefont{Hu et~al.}(2005)\citenamefont{Hu, Colgan, and
  Collins}}]{HuCoCo2005}
\bibinfo{author}{\bibfnamefont{S.~X.} \bibnamefont{Hu}},
  \bibinfo{author}{\bibfnamefont{J.}~\bibnamefont{Colgan}}, \bibnamefont{and}
  \bibinfo{author}{\bibfnamefont{L.~A.} \bibnamefont{Collins}},
  \bibinfo{journal}{J. Phys. B} \textbf{\bibinfo{volume}{38}},
  \bibinfo{pages}{L35} (\bibinfo{year}{2005}).

\bibitem[{\citenamefont{Foumouo et~al.}(2006)\citenamefont{Foumouo,
  Lagmago~Kamta, Edah, and Piraux}}]{FouLagEdaPir2006}
\bibinfo{author}{\bibfnamefont{E.}~\bibnamefont{Foumouo}},
  \bibinfo{author}{\bibfnamefont{G.}~\bibnamefont{Lagmago~Kamta}},
  \bibinfo{author}{\bibfnamefont{G.}~\bibnamefont{Edah}}, \bibnamefont{and}
  \bibinfo{author}{\bibfnamefont{B.}~\bibnamefont{Piraux}},
  \bibinfo{journal}{Phys. Rev. A} \textbf{\bibinfo{volume}{74}},
  \bibinfo{pages}{063409} (\bibinfo{year}{2006}).

\bibitem[{\citenamefont{Foumouo et~al.}(2008)\citenamefont{Foumouo, Antoine,
  Piraux, Malegat, Bachau, and Shakeshaft}}]{FouAntPir2008}
\bibinfo{author}{\bibfnamefont{E.}~\bibnamefont{Foumouo}},
  \bibinfo{author}{\bibfnamefont{P.}~\bibnamefont{Antoine}},
  \bibinfo{author}{\bibfnamefont{B.}~\bibnamefont{Piraux}},
  \bibinfo{author}{\bibfnamefont{L.}~\bibnamefont{Malegat}},
  \bibinfo{author}{\bibfnamefont{H.}~\bibnamefont{Bachau}}, \bibnamefont{and}
  \bibinfo{author}{\bibfnamefont{R.}~\bibnamefont{Shakeshaft}},
  \bibinfo{journal}{J. Phys. B} \textbf{\bibinfo{volume}{41}},
  \bibinfo{pages}{051001} (\bibinfo{year}{2008}).

\bibitem[{\citenamefont{Ivanov and Kheifets}(2007)}]{IvaKhe2007}
\bibinfo{author}{\bibfnamefont{I.~A.} \bibnamefont{Ivanov}} \bibnamefont{and}
  \bibinfo{author}{\bibfnamefont{A.~S.} \bibnamefont{Kheifets}},
  \bibinfo{journal}{Phys. Rev. A} \textbf{\bibinfo{volume}{75}},
  \bibinfo{pages}{033411} (\bibinfo{year}{2007}).

\bibitem[{\citenamefont{Nikolopoulos and Lambropoulos}(2007)}]{NikLam2007}
\bibinfo{author}{\bibfnamefont{L.~A.~A.} \bibnamefont{Nikolopoulos}}
  \bibnamefont{and}
  \bibinfo{author}{\bibfnamefont{P.}~\bibnamefont{Lambropoulos}},
  \bibinfo{journal}{J. Phys. B} \textbf{\bibinfo{volume}{40}},
  \bibinfo{pages}{1347} (\bibinfo{year}{2007}).

\bibitem[{\citenamefont{Nikolopoulos and Lambropoulos}(2001)}]{NikLam2001}
\bibinfo{author}{\bibfnamefont{L.~A.~A.} \bibnamefont{Nikolopoulos}}
  \bibnamefont{and}
  \bibinfo{author}{\bibfnamefont{P.}~\bibnamefont{Lambropoulos}},
  \bibinfo{journal}{J. Phys. B} \textbf{\bibinfo{volume}{34}},
  \bibinfo{pages}{545} (\bibinfo{year}{2001}).

\bibitem[{\citenamefont{Feng and van~der Hart}(2003)}]{FenHar2003}
\bibinfo{author}{\bibfnamefont{L.}~\bibnamefont{Feng}} \bibnamefont{and}
  \bibinfo{author}{\bibfnamefont{H.~W.} \bibnamefont{van~der Hart}},
  \bibinfo{journal}{J. Phys. B} \textbf{\bibinfo{volume}{36}},
  \bibinfo{pages}{L1} (\bibinfo{year}{2003}).

\bibitem[{\citenamefont{Horner et~al.}(2007)\citenamefont{Horner, Morales,
  Rescigno, Mart\'{i}n, and McCurdy}}]{HorMorRes2007}
\bibinfo{author}{\bibfnamefont{D.~A.} \bibnamefont{Horner}},
  \bibinfo{author}{\bibfnamefont{F.}~\bibnamefont{Morales}},
  \bibinfo{author}{\bibfnamefont{T.~N.} \bibnamefont{Rescigno}},
  \bibinfo{author}{\bibfnamefont{F.}~\bibnamefont{Mart\'{i}n}},
  \bibnamefont{and} \bibinfo{author}{\bibfnamefont{C.~W.}
  \bibnamefont{McCurdy}}, \bibinfo{journal}{Phys. Rev. A}
  \textbf{\bibinfo{volume}{76}}, \bibinfo{pages}{030701(R)}
  (\bibinfo{year}{2007}).

\bibitem[{\citenamefont{Hasegawa et~al.}(2005)\citenamefont{Hasegawa,
  Takahashi, Nabekawa, Ishikawa, and Midorikawa}}]{HasTakNabIsh2005}
\bibinfo{author}{\bibfnamefont{H.}~\bibnamefont{Hasegawa}},
  \bibinfo{author}{\bibfnamefont{E.~J.} \bibnamefont{Takahashi}},
  \bibinfo{author}{\bibfnamefont{Y.}~\bibnamefont{Nabekawa}},
  \bibinfo{author}{\bibfnamefont{K.~L.} \bibnamefont{Ishikawa}},
  \bibnamefont{and}
  \bibinfo{author}{\bibfnamefont{K.}~\bibnamefont{Midorikawa}},
  \bibinfo{journal}{Phys. Rev. A} \textbf{\bibinfo{volume}{71}},
  \bibinfo{pages}{023407} (\bibinfo{year}{2005}).

\bibitem[{\citenamefont{Nabekawa et~al.}(2005)\citenamefont{Nabekawa, Hasegawa,
  Takahashi, and Midorikawa}}]{NabHasTakMid2005}
\bibinfo{author}{\bibfnamefont{Y.}~\bibnamefont{Nabekawa}},
  \bibinfo{author}{\bibfnamefont{H.}~\bibnamefont{Hasegawa}},
  \bibinfo{author}{\bibfnamefont{E.~J.} \bibnamefont{Takahashi}},
  \bibnamefont{and}
  \bibinfo{author}{\bibfnamefont{K.}~\bibnamefont{Midorikawa}},
  \bibinfo{journal}{Phys. Rev. Lett.} \textbf{\bibinfo{volume}{94}},
  \bibinfo{pages}{043001} (\bibinfo{year}{2005}).

\bibitem[{\citenamefont{Sorokin et~al.}(2007)\citenamefont{Sorokin, Wellhofer,
  Bobashev, Tiedtke, and Richter}}]{SorWelBob2007}
\bibinfo{author}{\bibfnamefont{A.~A.} \bibnamefont{Sorokin}},
  \bibinfo{author}{\bibfnamefont{M.}~\bibnamefont{Wellhofer}},
  \bibinfo{author}{\bibfnamefont{S.~V.} \bibnamefont{Bobashev}},
  \bibinfo{author}{\bibfnamefont{K.}~\bibnamefont{Tiedtke}}, \bibnamefont{and}
  \bibinfo{author}{\bibfnamefont{M.}~\bibnamefont{Richter}},
  \bibinfo{journal}{Phys. Rev. A} \textbf{\bibinfo{volume}{75}},
  \bibinfo{pages}{051402(R)} (\bibinfo{year}{2007}).

\bibitem[{\citenamefont{Pindzola et~al.}(2007)\citenamefont{Pindzola,
  Robicheaux, Loch, Berengut, Topcu, Colgan, Foster, Griffin, Ballance, Schultz
  et~al.}}]{PinRobLoc2007}
\bibinfo{author}{\bibfnamefont{M.~S.} \bibnamefont{Pindzola}},
  \bibinfo{author}{\bibfnamefont{F.}~\bibnamefont{Robicheaux}},
  \bibinfo{author}{\bibfnamefont{S.~D.} \bibnamefont{Loch}},
  \bibinfo{author}{\bibfnamefont{J.~C.} \bibnamefont{Berengut}},
  \bibinfo{author}{\bibfnamefont{T.}~\bibnamefont{Topcu}},
  \bibinfo{author}{\bibfnamefont{J.}~\bibnamefont{Colgan}},
  \bibinfo{author}{\bibfnamefont{M.}~\bibnamefont{Foster}},
  \bibinfo{author}{\bibfnamefont{D.~C.} \bibnamefont{Griffin}},
  \bibinfo{author}{\bibfnamefont{C.~P.} \bibnamefont{Ballance}},
  \bibinfo{author}{\bibfnamefont{D.~R.} \bibnamefont{Schultz}},
  \bibnamefont{et~al.}, \bibinfo{journal}{J. Phys. B}
  \textbf{\bibinfo{volume}{40}}, \bibinfo{pages}{R39} (\bibinfo{year}{2007}).

\bibitem[{\citenamefont{Rescigno and McCurdy}(2000)}]{ResMcc2000}
\bibinfo{author}{\bibfnamefont{T.~N.} \bibnamefont{Rescigno}} \bibnamefont{and}
  \bibinfo{author}{\bibfnamefont{C.~W.} \bibnamefont{McCurdy}},
  \bibinfo{journal}{Phys. Rev. A} \textbf{\bibinfo{volume}{62}},
  \bibinfo{pages}{032706} (\bibinfo{year}{2000}).

\bibitem[{\citenamefont{McCurdy et~al.}(2001)\citenamefont{McCurdy, Horner, and
  Rescigno}}]{MccHorRes2001}
\bibinfo{author}{\bibfnamefont{C.~W.} \bibnamefont{McCurdy}},
  \bibinfo{author}{\bibfnamefont{D.~A.} \bibnamefont{Horner}},
  \bibnamefont{and} \bibinfo{author}{\bibfnamefont{T.~N.}
  \bibnamefont{Rescigno}}, \bibinfo{journal}{Phys. Rev. A}
  \textbf{\bibinfo{volume}{63}}, \bibinfo{pages}{022711}
  (\bibinfo{year}{2001}).

\bibitem[{\citenamefont{Schneider and Collins}(2005)}]{Schneider05}
\bibinfo{author}{\bibfnamefont{B.~I.} \bibnamefont{Schneider}}
  \bibnamefont{and} \bibinfo{author}{\bibfnamefont{L.~A.}
  \bibnamefont{Collins}}, \bibinfo{journal}{J. Non-Cryst. Solids}
  \textbf{\bibinfo{volume}{351}}, \bibinfo{pages}{1551} (\bibinfo{year}{2005}).

\bibitem[{\citenamefont{Schneider et~al.}(2006)\citenamefont{Schneider,
  Collins, and Hu}}]{SchColHu2006}
\bibinfo{author}{\bibfnamefont{B.~I.} \bibnamefont{Schneider}},
  \bibinfo{author}{\bibfnamefont{L.~A.} \bibnamefont{Collins}},
  \bibnamefont{and} \bibinfo{author}{\bibfnamefont{S.~X.} \bibnamefont{Hu}},
  \bibinfo{journal}{Phys. Rev. E} \textbf{\bibinfo{volume}{73}},
  \bibinfo{pages}{036708} (\bibinfo{year}{2006}).

\bibitem[{\citenamefont{McCurdy et~al.}(2004)\citenamefont{McCurdy, Baertschy,
  and Rescigno}}]{MccBaeRes2004}
\bibinfo{author}{\bibfnamefont{C.~W.} \bibnamefont{McCurdy}},
  \bibinfo{author}{\bibfnamefont{M.}~\bibnamefont{Baertschy}},
  \bibnamefont{and} \bibinfo{author}{\bibfnamefont{T.~N.}
  \bibnamefont{Rescigno}}, \bibinfo{journal}{J. Phys. B}
  \textbf{\bibinfo{volume}{37}}, \bibinfo{pages}{R137} (\bibinfo{year}{2004}).

\bibitem[{\citenamefont{McCurdy and Mart\'{i}n}(2004)}]{MccMar2004}
\bibinfo{author}{\bibfnamefont{C.~W.} \bibnamefont{McCurdy}} \bibnamefont{and}
  \bibinfo{author}{\bibfnamefont{F.}~\bibnamefont{Mart\'{i}n}},
  \bibinfo{journal}{J. Phys. B} \textbf{\bibinfo{volume}{37}},
  \bibinfo{pages}{917} (\bibinfo{year}{2004}).

\bibitem[{\citenamefont{Park and Light}(1986)}]{ParkLight86}
\bibinfo{author}{\bibfnamefont{T.~J.} \bibnamefont{Park}} \bibnamefont{and}
  \bibinfo{author}{\bibfnamefont{J.~C.} \bibnamefont{Light}},
  \bibinfo{journal}{J. Chem. Phys.} \textbf{\bibinfo{volume}{85}},
  \bibinfo{pages}{5870} (\bibinfo{year}{1986}).

\bibitem[{\citenamefont{Smyth et~al.}(1998)\citenamefont{Smyth, Parker, and
  Taylor}}]{SmyParTay1998}
\bibinfo{author}{\bibfnamefont{E.~S.} \bibnamefont{Smyth}},
  \bibinfo{author}{\bibfnamefont{J.~S.} \bibnamefont{Parker}},
  \bibnamefont{and} \bibinfo{author}{\bibfnamefont{K.~T.}
  \bibnamefont{Taylor}}, \bibinfo{journal}{Comput. Phys. Commun.}
  \textbf{\bibinfo{volume}{114}}, \bibinfo{pages}{1} (\bibinfo{year}{1998}).

\bibitem[{\citenamefont{Leforestier et~al.}(1991)\citenamefont{Leforestier,
  Bisseling, Cerjan, Feit, Friesner, Guldberg, Hammerich, Jolicard, Karrlein,
  Meyer et~al.}}]{Lefo90}
\bibinfo{author}{\bibfnamefont{C.}~\bibnamefont{Leforestier}},
  \bibinfo{author}{\bibfnamefont{R.~H.} \bibnamefont{Bisseling}},
  \bibinfo{author}{\bibfnamefont{C.}~\bibnamefont{Cerjan}},
  \bibinfo{author}{\bibfnamefont{M.~D.} \bibnamefont{Feit}},
  \bibinfo{author}{\bibfnamefont{R.}~\bibnamefont{Friesner}},
  \bibinfo{author}{\bibfnamefont{A.}~\bibnamefont{Guldberg}},
  \bibinfo{author}{\bibfnamefont{A.}~\bibnamefont{Hammerich}},
  \bibinfo{author}{\bibfnamefont{G.}~\bibnamefont{Jolicard}},
  \bibinfo{author}{\bibfnamefont{W.}~\bibnamefont{Karrlein}},
  \bibinfo{author}{\bibfnamefont{H.-D.} \bibnamefont{Meyer}},
  \bibnamefont{et~al.}, \bibinfo{journal}{J. Comp. Phys.}
  \textbf{\bibinfo{volume}{94}}, \bibinfo{pages}{59} (\bibinfo{year}{1991}).

\bibitem[{\citenamefont{Colgan et~al.}(2001)\citenamefont{Colgan, Pindzola, and
  Robicheaux}}]{ColPinRob2001}
\bibinfo{author}{\bibfnamefont{J.}~\bibnamefont{Colgan}},
  \bibinfo{author}{\bibfnamefont{M.~S.} \bibnamefont{Pindzola}},
  \bibnamefont{and}
  \bibinfo{author}{\bibfnamefont{F.}~\bibnamefont{Robicheaux}},
  \bibinfo{journal}{J. Phys. B} \textbf{\bibinfo{volume}{34}},
  \bibinfo{pages}{L457} (\bibinfo{year}{2001}).

\bibitem[{\citenamefont{Palacios et~al.}(2008)\citenamefont{Palacios, Rescigno,
  and McCurdy}}]{PalResMcc2008}
\bibinfo{author}{\bibfnamefont{A.}~\bibnamefont{Palacios}},
  \bibinfo{author}{\bibfnamefont{T.~N.} \bibnamefont{Rescigno}},
  \bibnamefont{and} \bibinfo{author}{\bibfnamefont{C.~W.}
  \bibnamefont{McCurdy}}, \bibinfo{journal}{Phys. Rev. A}
  \textbf{\bibinfo{volume}{77}}, \bibinfo{pages}{032716}
  (\bibinfo{year}{2008}).

\bibitem[{\citenamefont{Guan et~al.}(2008)\citenamefont{Guan, Bartschat, and
  Schneider}}]{GuaBarSch2008}
\bibinfo{author}{\bibfnamefont{X.}~\bibnamefont{Guan}},
  \bibinfo{author}{\bibfnamefont{K.}~\bibnamefont{Bartschat}},
  \bibnamefont{and} \bibinfo{author}{\bibfnamefont{B.~I.}
  \bibnamefont{Schneider}}, \bibinfo{journal}{Phys. Rev. A}
  \textbf{\bibinfo{volume}{77}}, \bibinfo{pages}{043421}
  (\bibinfo{year}{2008}).

\end{thebibliography}
\end{document}